\RequirePackage{etoolbox}
\csdef{input@path}{%
 {sty/}% cls, sty files
 {img/}% eps files
}%
\csgdef{bibdir}{bib/}% bst, bib files

\documentclass[ba]{imsart}
\usepackage{amsthm}
\usepackage{amsmath}
\usepackage{natbib}
\usepackage[colorlinks,citecolor=blue,urlcolor=blue,filecolor=blue,backref=page]{hyperref}
\usepackage{graphicx}
\RequirePackage[OT1]{fontenc}
\RequirePackage{amsthm,amsmath}
\RequirePackage{natbib}
\usepackage[utf8]{inputenc}
\usepackage{natbib}
\usepackage{amsthm}
\usepackage{amsfonts}
\usepackage{mathtools}
\usepackage{amssymb}
\usepackage{graphicx}
\usepackage{float}
\usepackage{pgfplotstable}
\usepackage{booktabs}
\usepackage{colortbl}
\usepackage{multirow}
\usepackage{array}
\usepackage{longtable}
\usepackage{tabu}
\usepackage{rotating}
\usepackage{lscape}
\usepackage{afterpage}
\usepackage{multirow}
\usepackage{subfigure}
\usepackage{listings}
\usepackage{color}
\usepackage{dsfont}
\usepackage{pdflscape}
\usepackage[left=4.5cm,right=4.5cm,top=2.5cm,bottom=2.5cm,footnotesep=0.5cm]{geometry}

\newcommand\norm[1]{\left\lVert#1\right\rVert}

\startlocaldefs
\endlocaldefs

\begin{document}

\begin{frontmatter}
\title{Variable selection in \\seemingly unrelated regressions\\ with random predictors}

\runtitle{Variable selection in SUR models}
%\thankstext{T1}{Footnote to the title with the ``thankstext'' command.}

\begin{aug}
\author{\fnms{David} \snm{Puelz}\ead[label=e1]{david.puelz@utexas.edu}},
\author{\fnms{P. Richard} \snm{Hahn}\ead[label=e2]{richard.hahn@chicagobooth.edu}}
\and
\author{\fnms{Carlos M.} \snm{Carvalho}
\ead[label=e3]{carlos.carvalho@mccombs.utexas.edu}}
\runauthor{Puelz, Hahn and Carvalho}

\affiliation{The University of Texas and The University of Chicago}

\address{David Puelz\\
\printead{e1}\\}

\address{P. Richard Hahn\\
\printead{e2}\\}

\address{Carlos M. Carvalho\\
\printead{e3}\\}

\end{aug}

\begin{abstract}
This paper considers linear model selection when the response is vector-valued and the predictors are randomly observed.  We  propose a new approach that decouples statistical inference from the selection step in a ``post-inference model summarization'' strategy.  We study the impact of predictor uncertainty on the model selection procedure.  The method is demonstrated through an application to asset pricing.
\end{abstract}

%\begin{keyword}[class=MSC]
%\kwd[Primary ]{60K35}
%\kwd{60K35}
%\kwd[; secondary ]{60K35}
%\end{keyword}
%
%\begin{keyword}
%\kwd{sample}
%\kwd{\LaTeXe}
%\end{keyword}

\end{frontmatter}

\section{Introduction and overview}
This paper develops a method for parsimoniously summarizing the shared dependence of many individual response variables upon a common set of predictor variables drawn at random. The focus is on multivariate Gaussian linear models where an analyst wants to find, among $p$ available predictors $X$, a subset which work well for predicting $q > 1$ response variables $Y$. The multivariate normal linear model assumes that a set of responses $\{ Y_{j} \}_{j=1}^{q}$ are linearly related to a shared set of covariates $\{ X_{i} \}_{i=1}^{p}$ via
\begin{equation}\label{modelfirst}
\begin{split}
	Y_{j} &= \beta_{j1}X_{1} + \cdots + \beta_{jp}X_{p} + \epsilon_{j}, \;\;\;\;\; \boldsymbol{\epsilon} \sim \mbox{N}(0, \Psi),
\end{split}
\end{equation}
where $\Psi$ is a non-diagonal covariance matrix. 

Bayesian variable selection in (single-response) linear models is the subject of a vast literature, from prior specification on parameters \citep{Berger12} and models \citep{ScottBerger06} to efficient search strategies over the model space \citep{GeorgeandMcCulloch, hans2007shotgun}. For a more complete set of references we refer the reader to the reviews of \cite{Clyde04} and \cite{HahnCarvalho}. By comparison, variable selection has not been widely studied in concurrent regression models, perhaps because it is natural simply to apply existing variable selection methods to each univariate regression individually. Indeed, such joint regression models go by the name ``seemingly unrelated regressions'' (SUR) in the Bayesian econometrics literature, reflecting the fact that the regression coefficients from each of the separate regressions can be obtained in isolation from one another (i.e., conducting estimation as if $\Psi$ were diagonal). However, allowing non-diagonal $\Psi$ can lead to more efficient estimation \citep{zellner1962efficient} and can similarly impact variable selection \citep{brown1998multivariate, wangSUR}.

%The unknown parameters $\Theta = \lbrace \boldsymbol \beta, \Psi \rbrace$ can be inferred via Bayesian conditioning. 

This paper differs from \cite{brown1998multivariate} and \cite{wangSUR} in that we focus on the case where the predictor variables (the regressors, or covariates) are treated as random as opposed to fixed.Our goal will be to summarize codependence among multiple responses in {\em subsequent} periods, making the uncertainty in future realizations highly central to our selection objective.  This approach is natural in many contexts (e.g., macroeconomic models) where the purpose of selection is inherently forward-looking. To our knowledge, no existing variable selection methods are suitable in this context. The new approach is based on the sparse summary perspective outlined in \cite{HahnCarvalho}, which applies Bayesian decision theory to summarize complex posterior distributions. By using a utility function that explicitly rewards sparse summaries, a high dimensional posterior distribution is collapsed into a more interpretable sequence of sparse point summaries.

A related approach to variable selection in multivariate Gaussian models is the Gaussian graphical model framework \citep{jones2005experiments,dobra2004sparse,wang2009bayesian}. In that approach, the full conditional distributions are represented in terms of a sparse $(p+q)$-by-$(p+q)$ precision matrix. By contrast, we partition the model into response and predictor variable blocks, leading to a distinct selection criterion that narrowly considers the $p$-by-$q$ covariance between $Y$ to $X$.
% see the definitions of $S$ and $A$ in (\ref{moments}). 

\subsection{Methods overview}\label{overview} Posterior summary variable selection consists of three phases: {\em model specification and fitting}, {\em utility specification}, and {\em graphical summary}. Each of these steps is outlined below. Additional details of the implementation are described in Section \ref{DSS} and the Appendix.

\subsubsection*{Step 1: Model specification and fitting}
The statistical model may be described compositionally as $p(Y,X) = p(Y \vert X)p(X)$. For $(Y,X) \sim \mbox{N}(\mu,\Sigma)$, the regression model  (\ref{modelfirst}) implies $\Sigma$ has the following block structure:

\begin{align}\label{model2}
\Sigma
= 
\left[
\begin{array}{c|c}
{\boldsymbol \beta}^{T}\Sigma_{x}{\boldsymbol\beta} + \Psi & (\Sigma_{x}{\boldsymbol\beta})^{T} \\
\hline
\Sigma_{x}{\boldsymbol\beta}
 & 
\Sigma_{x} \\
\end{array}
\right].
\end{align}
We denote the unknown parameters for the full joint model as $\Theta = \{\mu_{x},\mu_{y},\Sigma_{x},\boldsymbol{\beta},\Psi\}$ where $\mu = (\mu_y^T, \mu_x^T)^T$ and $\Sigma_x = \mbox{cov}(X)$.  

For a given prior choice $p(\Theta)$, posterior samples of all model parameters are computed by routine Monte Carlo methods, primarily Gibbs sampling. Details of the specific modeling choices and associated posterior sampling strategies are described in the Appendix. 

A notable feature of our approach is that {\it steps 2} (and {\it 3}) will be unaffected by modeling choices made in {\it step 1} except insofar as they lead to different posterior distributions $p(\Theta \vert \mathbf{Y}, \mathbf{X})$. In short, {\it step 1} is ``obtain a posterior distribution''; posterior samples then become inputs to {\it step 2}.

\subsubsection*{Step 2: Utility specification}
For our utility function we use the log-density of the regression $p(Y \vert X)$ above. It is convenient to work in terms of negative utility, or loss:
\begin{equation}
	\begin{split}
		\mathcal{L}(\tilde{Y},\tilde{X},\Theta,\boldsymbol{\gamma}) = \frac{1}{2}( \tilde{Y} - \boldsymbol{\gamma}\tilde{X} )^{T} \Omega ( \tilde{Y} - \boldsymbol{\gamma}\tilde{X} ),
	\end{split}
\end{equation}
 where $\Omega = \Psi^{-1}$. Note that this log-density is being used in a descriptive capacity, not an inferential one; that is, all posterior inferences are based on the posterior distribution from {\it step 1}. The ``action'' $\boldsymbol{\gamma}$ is regarded as a point estimate of the regression parameters $\boldsymbol{\beta}$, which would be a good fit to {\em future} data $(\tilde{Y}, \tilde{X})$ drawn from the same model as the observed data. 
 
 Taking expectations over the posterior distribution of all unknowns
 \begin{equation}
	\begin{split}
		p(\tilde{Y},\tilde{X}, \Theta \vert \textbf{Y}, \textbf{X}) =  p(\tilde{Y} \vert \tilde{X}, \Theta) p(\tilde{X} \vert \Theta) p(\Theta \vert \textbf{Y}, \textbf{X}),
	\end{split}
\end{equation}
 yields expected loss
 \begin{equation}
	\mathcal{L}(\boldsymbol{\gamma}) \equiv \mathbb{E}[ \mathcal{L}(\tilde{Y},\tilde{X}, \Theta, \boldsymbol{\gamma}) ] =\text{tr}[ M \boldsymbol{\gamma} S \boldsymbol{\gamma}^{T} ] - 2\text{tr}[A\boldsymbol{\gamma}^{T}] + \mbox{constant},
\end{equation}where $A=\mathbb{E}[\Omega\tilde{Y}\tilde{X}^{T}]$, $S=\mathbb{E}[\tilde{X}\tilde{X}^{T}] = \overline{\Sigma_{x}}$, and $M=\overline{\Omega}$, the overlines denote posterior means, and the final term is a constant with respect to $\boldsymbol{\gamma}$.

Finally, we add an explicit penalty, reflecting our preference for sparse summaries: 
%will result in optimal summaries that have exactly zero coefficients:
 \begin{equation}\label{ex_loss}
	\mathcal{L}_{\lambda}(\boldsymbol{\gamma}) \equiv \text{tr}[ M \boldsymbol{\gamma} S \boldsymbol{\gamma}^{T} ] - 2\text{tr}[A\boldsymbol{\gamma}^{T}] + \lambda \norm{\text{{\bf vec}}(\boldsymbol{\gamma})}_{0},
\end{equation}
 where $\norm{\text{{\bf vec}}(\boldsymbol{\gamma})}_{0}$ counts the number of non-zero elements in $\boldsymbol{\gamma}$. In practice, we will use an approximation to this utility based on the $\ell_1$ penalty; optimal actions under this approximation will still be sparse.
\subsubsection*{Step 3: Graphical summary}
Traditional applications of Bayesian decision theory derive {\em point-estimates} by minimizing expected loss for certain loss functions. The present goal is not an {\em estimator} per se, but a parsimonious summary of information contained in a complicated, high dimensional posterior distribution. This distinction is worth emphasizing because we have not one, but rather a continuum of loss functions, indexed by the penalty parameter $\lambda$. This class of loss functions can be used to organize the posterior distribution as follows.  

Using available convex optimization techniques, expression (\ref{ex_loss}) can be optimized efficiently for a range of $\lambda$ values simultaneously. Posterior graphical summaries consist of two components. First, graphs depicting which response variables have non-zero $\boldsymbol{\gamma}_{\lambda}^*$ coefficients on which predictor variables can be produced for any given $\lambda$. Second, posterior distributions of the quantity
\begin{equation}
\Delta_{\lambda} = \mathcal{L}(\tilde{Y},\tilde{X},\Theta,\boldsymbol{\gamma}_{\lambda}^*) - \mathcal{L}(\tilde{Y},\tilde{X},\Theta,\boldsymbol{\gamma}^*) 
\end{equation}
can be used to gauge the impact $\lambda$ has on the descriptive capacity of $\gamma_{\lambda}^*$.  Here, $\boldsymbol{\gamma}^* = \gamma_{\lambda=0}^*$ is the unpenalized optimal solution to the minimization of loss (\ref{ex_loss}). %The more posterior mass of $\Delta_{\lambda}$ piled about 0, the higher fidelity the corresponding posterior summary, $\gamma_{\lambda}^*$. 

%In practice, it is often found that substantial penalization (leading to sparse posterior graphs) does not dramatically disperse the corresponding distribution of $\Delta_{\lambda}$ \textcolor{red}{\it We've actually now seen that it does, but that's okay}.

%Conditioning on data $(\mathbf{Y},\mathbf{X})$, we obtain posterior samples of the joint model's parameters from the posterior distribution, $p(\Theta \vert \mathbf{Y},\mathbf{X})$.

\section{Posterior summary variable selection}\label{DSS}
The statistical model is given in equations (\ref{modelfirst}) and (\ref{model2}); prior specification and model fitting details can be found in the Appendix. Alternatively, the models described in \cite{brown1998multivariate} or \cite{wangSUR} could be used. In this section, we flesh out the details of {\it steps 2} and {\it 3}, which represent the main contributions of this paper.

%\subsection{Utility-based variable selection}
%Assume we have obtained a posterior distribution for the model parameters: $p(\Theta \vert \mathbf{Y},\mathbf{X})$. Our goal now is to describe the relationship between the responses and covariates.  While the parameters of our model do precisely this, after appropriate Bayesian conditioning, we only have posterior samples of these parameters rather than a simple point estimate. Moreover, these parameters are potentially ``larger" than we would like, in the sense that they involve all possible covariates while perhaps a much smaller number accounts for the vast majority of the covariance structure of the response vector. 
\subsection{Deriving the sparsifying expected utility function}
Define the optimal posterior summary as the $\boldsymbol{\gamma}^*$ minimizing some expected loss $\mathcal{L}_{\lambda}(\boldsymbol{\gamma}) = \mathbb{E}[\mathcal{L}_{\lambda}(\tilde{Y},\tilde{X},\Theta,\boldsymbol{\gamma})]$.  Here, the expectation is taken over the joint posterior predictive and posterior distribution: $p(\tilde{Y},\tilde{X}, \Theta \mid \textbf{Y}, \textbf{X})$. 

As described in the previous section, our loss takes the form of a penalized log conditional distribution:
\begin{equation}
	\begin{split}
		\mathcal{L}_{\lambda}(\tilde{Y},\tilde{X},\Theta,\boldsymbol{\gamma}) \equiv \frac{1}{2}( \tilde{Y} - \boldsymbol{\gamma}\tilde{X} )^{T} \Omega ( \tilde{Y} - \boldsymbol{\gamma}\tilde{X} )  + \lambda \norm{\text{{\bf vec}}(\boldsymbol{\gamma})}_{0}, \label{newlossstoch}
	\end{split}
\end{equation}where $\Omega = \Psi^{-1}$, $\norm{\text{{\bf vec}}(\boldsymbol{\gamma})}_{0} = \sum_{j}\mathds{1}\left(\text{{\bf vec}}(\boldsymbol{\gamma}) \neq 0\right)$, and $\text{{\bf vec}}(\boldsymbol{\gamma})$ is the vectorization of the action matrix $\boldsymbol{\gamma}.$ The first term of this loss measures the distance (weighted by the precision $\Omega$) between the linear predictor $\boldsymbol{\gamma}\tilde{X}$ and a future response $\tilde{Y}$.  The second term promotes a sparse optimal summary, $\boldsymbol{\gamma}$. The penalty parameter $\lambda$ determines  the relative importance of these two components.  Expanding the quadratic form gives:
\small
\begin{equation}\label{modnew1}
	\begin{split}
		\mathcal{L}_{\lambda}(\tilde{Y},\tilde{X}, \Theta, \boldsymbol{\gamma}) &= \frac{1}{2}\left(\tilde{Y}^{T} \Omega \tilde{Y} - 2\tilde{X}^{T}\boldsymbol{\gamma}^{T} \Omega \tilde{Y} + \tilde{X}^{T}\boldsymbol{\gamma}^{T}  \Omega \boldsymbol{\gamma} \tilde{X}\right) + \lambda \norm{\text{{\bf vec}}(\boldsymbol{\gamma})}_{0} \\
		&= \left(  \tilde{X}^{T}\boldsymbol{\gamma}^{T}  \Omega \boldsymbol{\gamma} \tilde{X} -2\tilde{X}^{T}\boldsymbol{\gamma}^{T} \Omega \tilde{Y}\right) + \lambda \norm{\text{{\bf vec}}(\boldsymbol{\gamma})}_{0} + \mbox{constant}.	
		\end{split}
\end{equation}
\normalsize
Integrating over $(\tilde{Y},\tilde{X},\Theta \mid \textbf{Y}, \textbf{X})$ (and dropping the constant) gives:  

%In this we take advantage of the following properties of the matrix trace: $\text{tr}[x] = x$ for scalar $x$ and $\text{tr}[ABC] = \text{tr}[CAB] = \text{tr}[BCA]$ for matrices $A$, $B$, and $C$.

\begin{equation}\label{almostlassoform}
	\begin{split}
		\mathcal{L}_{\lambda}(\boldsymbol{\gamma}) &= \mathbb{E}[ \mathcal{L}_{\lambda}(\tilde{Y},\tilde{X}, \Theta, \boldsymbol{\gamma}) ]\\
%		 &= \mathbb{E}\left[ \tilde{X}^{T}\boldsymbol{\gamma}^{T}  \Omega \boldsymbol{\gamma} \tilde{X} \right] - \mathbb{E}\left[ 2\tilde{X}^{T}\boldsymbol{\gamma}^{T} \Omega \tilde{Y} \right] + \lambda \norm{\text{ {\bf vec}}(\boldsymbol{\gamma})}_{0}\\
		 &= \mathbb{E}\left[ \text{tr}[ \boldsymbol{\gamma}^{T}  \Omega \boldsymbol{\gamma} \tilde{X} \tilde{X}^{T}] \right] - 2\mathbb{E}\left[ \text{tr}[\boldsymbol{\gamma}^{T} \Omega \tilde{Y}\tilde{X}^{T}] \right] + \lambda \norm{\text{{\bf vec}}(\boldsymbol{\gamma})}_{0}, \\
		&=  \mathbb{E}\left[ \text{tr}[ \boldsymbol{\gamma}^{T}  \Omega \boldsymbol{\gamma} S] \right] - 2\text{tr}[A\boldsymbol{\gamma}^{T}] + \lambda \norm{\text{{\bf vec}}(\boldsymbol{\gamma})}_{0},\\
%		&= 2\text{tr}[\boldsymbol{\gamma}^{T}A] - \mathbb{E}\left[ \text{tr}[  \Omega \boldsymbol{\gamma} S \boldsymbol{\gamma}^{T} ] \right] + \lambda \norm{\text{ {\bf vec}}(\boldsymbol{\gamma})}_{0}\\
		&= \text{tr}[ M \boldsymbol{\gamma} S \boldsymbol{\gamma}^{T} ] - 2\text{tr}[A\boldsymbol{\gamma}^{T}]  + \lambda \norm{\text{{\bf vec}}(\boldsymbol{\gamma})}_{0},
	\end{split}
\end{equation}where 
\begin{equation}\label{moments}
\begin{split}
A &\equiv\mathbb{E}[\Omega\tilde{Y}\tilde{X}^{T}],\\
S&\equiv\mathbb{E}[\tilde{X}\tilde{X}^{T}] = \overline{\Sigma}_{x},\\
M&\equiv\overline{\Omega},
\end{split}
\end{equation}
and the overlines denote posterior means.  Define the Cholesky decompositions $M = LL^{T}$ and $S = QQ^{T}$. To make the optimization problem tractable we replace the $\ell_0$ norm with the $\ell_1$ norm, leading to an expression that can be formulated in the form of a standard penalized regression problem:
\begin{equation}\label{lasso_form}
	\mathcal{L}_{\lambda}(\boldsymbol{\gamma}) =  \norm{ \left[Q^{T} \otimes L^{T}\right]\text{\bf vec}(\boldsymbol{\gamma}) - \text{\bf vec}(L^{-1}AQ^{-T})  }_{2}^{2} + \lambda\norm{ \text{\bf vec}(\boldsymbol{\gamma})}_{1},
\end{equation}
with covariates $Q^{T} \otimes L^{T}$, ``data" $L^{-1}AQ^{-T}$, and regression coefficients $\boldsymbol{\gamma}$  (see the Appendix for details). Accordingly, (\ref{lasso_form}) can be optimized using existing software such as the {\tt lars} R package of \cite{Efron} and still yield sparse solutions.  %Equation \ref{lasso_form} represents completion of {\it step 2} in the posterior summary variable selection procedure.  The use of the $\ell_1$ norm defines a convex optimization problem that can be efficiently solved. This choice satisfies our goal of obtaining a solution path that can be evaluated relative to the unpenalized loss. 

\subsection{Sparsity-utility trade-off plots}
Rather than attempting to determine an ``optimal'' value of $\lambda$, we advocate displaying plots that reflect the utility attenuation due to $\lambda$-induced sparsification. We define the ``loss gap'' between a $\lambda$-sparse solution, $\mathcal{L}(\tilde{Y},\tilde{X},\Theta,\boldsymbol{\gamma}_{\lambda}^*)$, and the optimal unpenalized (non-sparse, $\lambda = 0$) summary, $\mathcal{L}(\tilde{Y},\tilde{X},\Theta,\boldsymbol{\gamma}^*)$ as
\begin{equation}
\Delta_{\lambda} = \mathcal{L}(\tilde{Y},\tilde{X},\Theta,\boldsymbol{\gamma}_{\lambda}^*) - \mathcal{L}(\tilde{Y},\tilde{X},\Theta,\boldsymbol{\gamma}^*). 
\end{equation}
As a function of $(\tilde{Y},\tilde{X}, \Theta)$, $\Delta_{\lambda}$ is itself a random variable which we can sample by obtaining posterior draws from $p(\tilde{Y},\tilde{X}, \Theta \mid \textbf{Y}, \textbf{X})$. The posterior distribution(s) of $\Delta_{\lambda}$ (for various $\lambda$) therefore reflects the deterioration in utility attributable to ``sparsification''. Plotting these distributions as a function of $\lambda$ allows one to visualize this trade-off. Specifically, $\pi_{\lambda} \equiv \mbox{Pr}(\Delta_{\lambda} < 0 \mid  \textbf{Y}, \textbf{X})$ is the (posterior) probability that the $\lambda$-sparse summary is no worse than the non-sparse summary. Using this framework, a useful heuristic for obtaining a single sparse summary is to report the sparsest model (associated with the highest $\lambda$) such that $\pi_{\lambda}$ is higher than some pre-determined threshold, $\kappa$; we adopt this approach in our application section.

We propose summarizing the posterior distribution of $\Delta_{\lambda}$ via two types of plots. First, one can examine posterior means and credible intervals of $\Delta_{\lambda}$ for a sequence of models indexed by $\lambda$. Similarly, one can plot $\pi_{\lambda}$ across the same sequence of models. Also, for a fixed value of $\lambda$, one can produce graphs where nodes represent predictor variables and response variables and an edge is drawn between nodes whenever the corresponding element of $\gamma^*_{\lambda}$ is non-zero. All three types of plots are exhibited in Section \ref{apps}.

\subsection{Relation to previous methods}

Loss function (\ref{lasso_form}) is similar in form to the univariate \textit{DSS} (decoupled shrinkage and selection) strategy developed by \cite{HahnCarvalho}. Our approach generalizes \cite{HahnCarvalho} by optimizing over the matrix $\boldsymbol{\gamma} \in \mathbb{R}^{qxp}$ rather than a single vector of regression coefficients, extending the sparse summary utility approach to seemingly unrelated regression models \citep{brown1998multivariate, wangSUR}. 

Additionally, the present method considers random predictors, $\tilde{X}$, whereas \cite{HahnCarvalho} considered only a matrix of fixed design points. The impact of accounting for random predictors on the posterior summary variable selection procedure is examined in more detail in the application section.

%This approach differs from the group lasso of \cite{grouplasso} (where grouped covariates enter the model simultaneously along the lasso solution path) in that {\it (i)} it does not impose a covariate grouping structure and {\it (ii)} the loss function presented is derived by integrating over uncertainty described by the joint distribution of all unknowns $(\tilde{Y}, \tilde{X}, \Theta)$. 

%In an appendix, \cite{HahnCarvalho} remark that their sparse utility summary approach can be applied to Gaussian graphical models and derive an expected utility that is equivalent to the graphical lasso \citep{Friedmanetal2008} applied to posterior moments. In contrast to this approach, our expected utility depends only on the off-diagonal block of the joint precision matrix rather than the full joint distribution are penalized choice variables. 
%We demonstrate the difference between the graphical lasso and our multivariate loss function (assuming random predictors) via a simple  example in the Appendix.

An important difference between the sparse summary utility approach and previous approaches is in the role played by the posterior distribution. Many Bayesian variable importance metrics are based on the posterior distribution of an indicator variable that records if a given variable is non-zero (included in the model). The model we will use in our application utilizes such an indicator vector, which is called $\alpha$.
%Importantly, this sampling procedure includes drawing a binary vector $\alpha$ of length $p$ specifying predictor $i$'s inclusion ($\alpha_{i}=1$) or exclusion ($\alpha_{i}=0$) from a sampled model.  Among the other parameters, we also sample a full residual covariance $\Psi$ to capture the joint dependence of each response on each predictor in selection.  
%Given posterior sample of the binary vector $\alpha$, it may be tempting to use it to directly select a model.  
For example, a widely-used model selection heuristic is to examine the ``inclusion probability'' of predictor $i$, defined as the posterior mean of component $\alpha_{i}$. However, any approach based on the posterior mean of $\alpha$ necessarily ignores information about the codependence between its elements, which can be substantial in cases of collinear predictors. Our method focuses instead on the expected log-density of future predictions, which synthesizes information from all parameters simultaneously in gauging how important they are in terms of future predictions.

\section{Applications}\label{apps}

In this section, the sparse posterior summary method is applied to a data set from the finance (asset pricing) literature. A key component of our analysis will be a comparison between the posterior summaries obtained when the predictors are drawn at random versus when they are assumed fixed.

The response variables are returns on 25 tradable portfolios and our predictor variables are returns on 10 other portfolios thought to be of theoretical importance. In the asset pricing literature \cite{Ross}, the response portfolios represent assets to be priced (so-called {\em test assets}) and the predictor portfolios represent distinct sources of variation (so-called {\em risk factors}). More specifically, the test assets $Y$ represent trading strategies based on company size (total value of stock shares) and book-to-market (the ratio of the company's accounting valuation to its size); see  \cite{FF3} and \cite{FF5} for details.  Roughly, these assets serve as a lower-dimensional proxy for the stock market. The risk factors are also portfolios, but ones which are thought to represent {\em distinct} sources of risk. What constitutes a distinct source of risk is widely debated, and many such factors have been proposed in the literature \citep{cochrane2011presidential}. We use monthly data from July 1963 through February 2015, obtained from Ken French's website:
\begin{center}
 {\tt http://mba.tuck.dartmouth.edu/pages/faculty/ken.french/}.
\end{center}

Our analysis investigates which subset of risk factors are most relevant (as defined by our utility function). As our initial candidates, we consider factors known in previous literature as: market, size, value,  direct profitability, investment, momentum, short term reversal, long term reversal, betting against beta, and quality minus junk.   Each factor is constructed by cross-sectionally sorting stocks by various characteristics of a company and forming linear combinations based on these sorts. For example, the value factor is constructed using the book-to-market ratio of a company.  A high ratio indicates the company's stock is a ``value stock" while a low ratio leads to a ``growth stock" assessment.  Essentially, the value factor is a portfolio built by going long stocks with high book-to-market ratio and shorting stocks with low book-to-market ratio.  For detailed definitions of the first five factors, see \cite{FF5}. In the figures to follow, each is labeled as, for example, ``Size2 BM3," to denote the portfolio buying stocks in the second quintile of size and the third quintile of book-to-market ratio.

%In these examples, we are interested in selecting risk factors for an asset pricing model where financial asset returns (often called test assets) are modeled as linear combinations of a set of risk factor returns.  In our general notation, the risk factors are $X$ and the test assets are $Y$.  It is postulated that exposure to these risk factors with distinct risk premia\footnote{return derived from the factor's source of risk} comprise the returns of the overall asset.  One motivating theory used extensively in this context is the arbitrage pricing theory (APT) of \cite{Ross}.  
%Although APT proposes a linear relationship between financial asset and risk factor returns, it does not mention which factors should be used.  Thus, several hundred potential factors, the so called ``factor zoo", have been proposed in the finance literature with little indication of which ones may be the \textit{true} factors \citep[see John Cochrane's discussion in][]{cochrane2011presidential}.  We model sets of test asset returns as linear combinations of potential risk factors in an attempt to identify which factors may be the most important for pricing.  
Recent related work includes \cite{ericsson2004choosing} and \cite{harvey2015lucky}. \cite{ericsson2004choosing} follow a Bayesian model selection approach based off of inclusion probabilities, representing the preliminary inference step of our methodology. \cite{harvey2015lucky} take a different approach that utilizes multiple hypothesis testing and bootstrapping.

% It is widely believed that these factors (or some subset of them) reflect dimensions of independent variation in the financial markets that investors are compensated for exposure to as discussed in \cite{FF3,FF5}.  

%Section \ref{DSS} completed {\it step 2} of our procedure: {\it utility specification}. 

\subsection{Results}
As described in Section \ref{overview}, the first step of our analysis consists of fitting a Bayesian model. We fit model (\ref{modelfirst}) using a variation of the well-known stochastic search variable selection algorithm of \cite{GeorgeandMcCulloch} and similar to \cite{brown1998multivariate} and \cite{wangSUR}. Details are given in the Appendix.  

In the subsections to follow, we will show the following two figures.  First, we plot the expectation of $\Delta_{\lambda}$ (and associated posterior credible interval) across a range of $\lambda$ penalties. Recall, $\Delta_{\lambda}$  is the ``loss gap'' between a sparse summary and the best non-sparse (saturated) summary, meaning that smaller values are ``better''. Additionally, we plot the probability that a given model is no worse than the saturated model $\pi_{\lambda}$ on this same figure, where ``no worse'' means $\Delta_{\lambda} < 0$. Note that even for very weak penalties (small $\lambda$), the distribution of $\Delta_{\lambda}$ will have non-zero variance and therefore even if it is centered about zero, some mass can be expected to fall above zero; practically, this means that $\pi_{\lambda} > 0.5$ is a very high score.

Second, we display a summary graph of the selected variables for the $\kappa=12.5\%$ threshold. Recall that this is the highest penalty (sparsest graph) that is no worse than the saturated model with $12.5\%$ posterior probability. For these graphs,  the response and predictor variables are colored gray and white, respectively. A test asset label of, for example, ``Size2 BM3," denotes the portfolio that buys stocks in the second quintile of size and the third quintile of book-to-market ratio. The predictors without connections to the responses under the optimal graph are not displayed. 
\newpage
These two figures are shown in four scenarios:
\begin{enumerate}
	\item Random predictors.
	\item Fixed predictors.
	\item Random predictors under alternative prior.
	\item Fixed predictors under alternative prior.
\end{enumerate}The ``alternative prior" scenario serves to show the impact of the statistical modeling comprising {\it step 1}. Specifically, we use the same Monte Carlo model fitting procedure as before (described in the Appendix) but fix $\alpha$ to the identity vector. That is, we omit the point-mass component of the priors for the elements of $\boldsymbol{\beta}$.

% is a different inference procedure where we fix $\alpha_{i}=1 \hspace{2mm} \forall i$ in our matrix-variate MCMC ($\alpha$ is the binary vector specifying a particular linear model, $M_{\alpha}$). This amounts to independently sampling the regression parameters of each test asset on \textbf{all} 10 factors.  In the first two scenarios, we also show a set of selected graphs for varying levels of the threshold parameter $\kappa$.

\subsubsection{Random predictors}

\begin{figure}[H]
\centering
\includegraphics[scale=0.47]{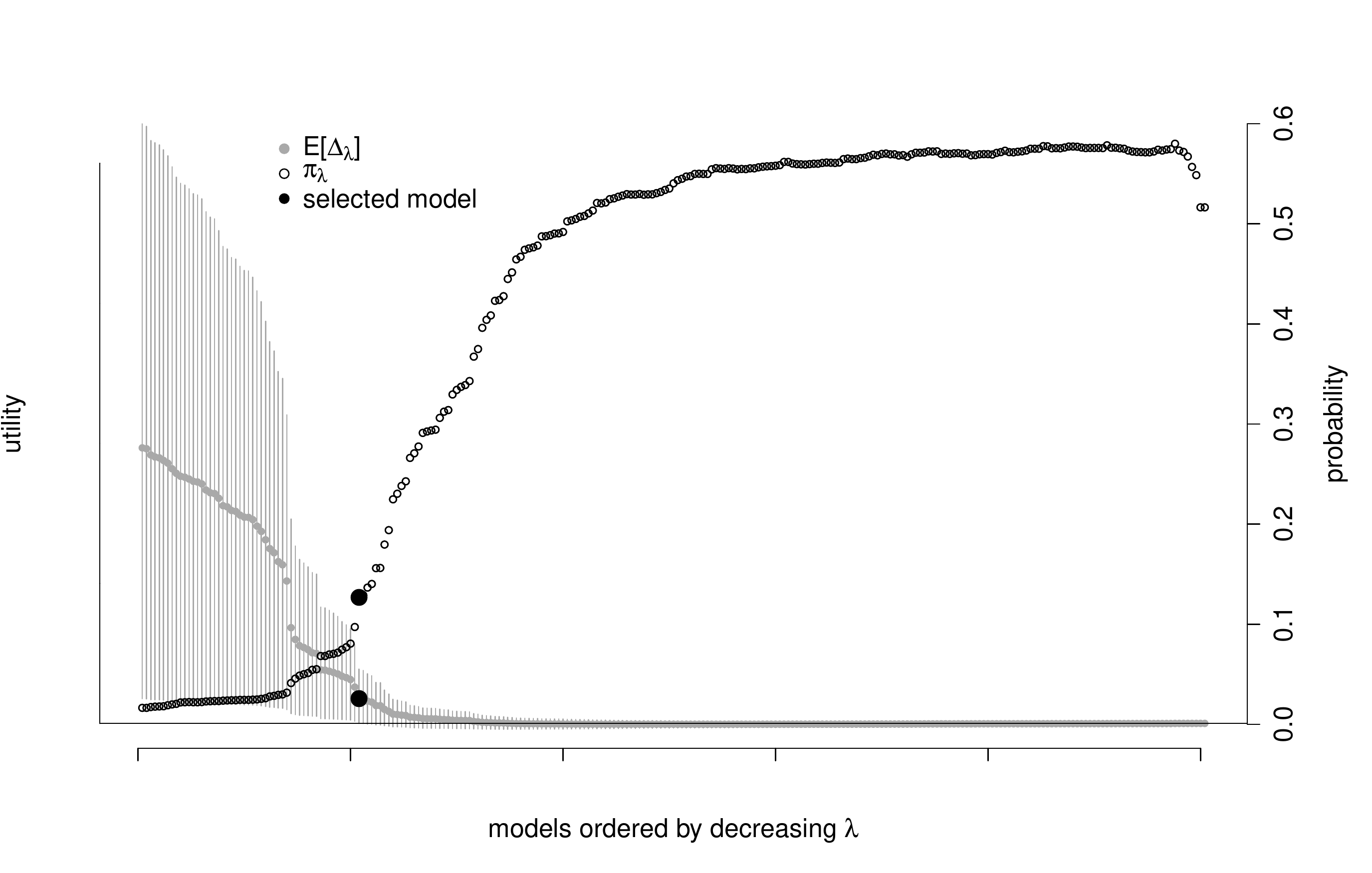}
  \caption{Evaluation of $\Delta_{\lambda}$ and $\pi_{\lambda}$ along the solution path for the 25 size/value portfolios modeled by the 10 factors. An analyst may use this plot to select a particular model. Uncertainty bands are 75\% posterior intervals on the $\Delta_{\lambda}$ metric. The large black dot represents the model selected in Figure \ref{graph25}.}
  \label{Lossgraph25}
   \includegraphics[scale=.65]{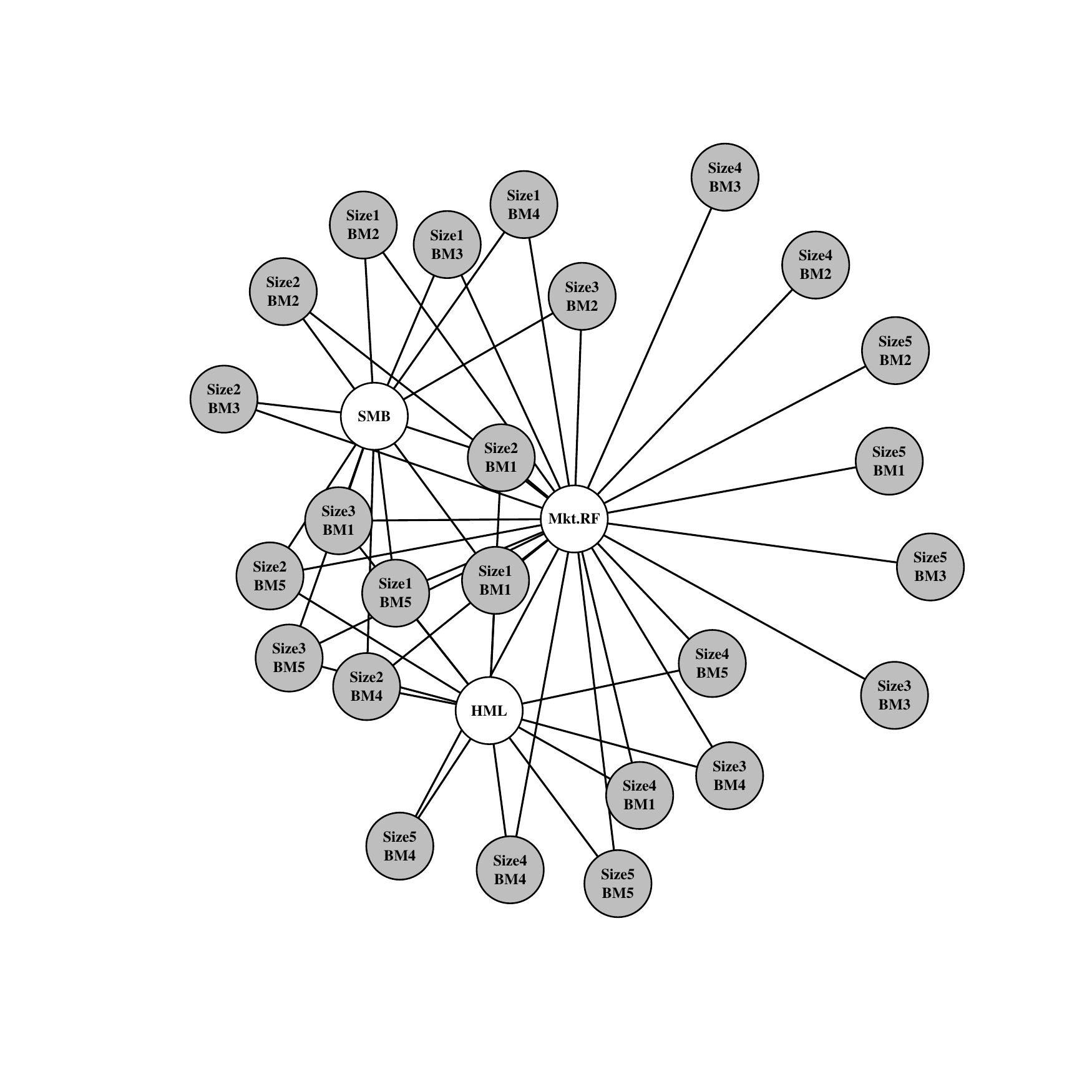}
   \vspace{-10mm}
  \caption{The selected model for 25 size/value portfolios modeled by the 10 factors.  The responses and predictors are colored in gray and white, respectively.  Edges represent nonzero components of the optimal action, $\boldsymbol{\gamma}$.}\label{graph25}
\end{figure}

This section introduces our baseline example where the risk factors (predictors) are random.  We evaluate the set of potential models by analyzing plots such as figure \ref{Lossgraph25}.  This shows $\Delta_{\lambda}$ and $\pi_{\lambda}$ evaluated across a range of $\lambda$ values.  Additionally, we display the posterior uncertainty in the $\Delta_{\lambda}$ metric with gray vertical uncertainty bands: these are the centered $P\%$ posterior credible intervals where $\kappa = (1-P)/2$.   As the accuracy of the sparsified solution increases, the posterior of $\Delta_{\lambda}$ concentrates around zero by construction, and the probability of the model being no worse than the saturated model, $\pi_{\lambda}$, increases.  We choose the sparsest model such that its corresponding $\pi_{\lambda} > \kappa = 12.5\%$.  This model is displayed in figure \ref{graph25} and is identified by the black dot in figure \ref{Lossgraph25}.

The selected set of factors in graph \ref{graph25} are the market (Mkt.RF), value (HML), and size (SMB). This three factor model is no worse than the saturated model with $12.5\%$ posterior probability where all test assets are connected to all risk factors.  Note also that in our selected model almost every test asset is distinctly tied to one of either value or size and the market factor.  These are the three factors of Ken French and Eugene Fama's pricing model developed in \cite{FF3}.  They are known throughout the finance community as being ``fundamental dimensions" of the financial market, and our procedure is consistent with this widely held belief at a small $\kappa$ level.  

The characteristics of the test assets in graph \ref{graph25} are also important to highlight. The test portfolios that invest in small companies (``Size1" and ``Size2") are primarily connected to the SMB factor which is designed as a proxy for the risk of small companies. Similarly, the test portfolios that invest in high book-to-market companies (``BM4" and ``BM5") have connections to the HML factor which is built on the idea that companies whose book value exceeds the market's perceived value should generate a distinct source of risk.  As previously noted, all of the test portfolios are connected to the market factor suggesting that it is a relevant predictor even for the sparse $\kappa=12.5\%$ selection criterion.
 
\begin{figure}[H]
\centerline{
\includegraphics[scale=.8]{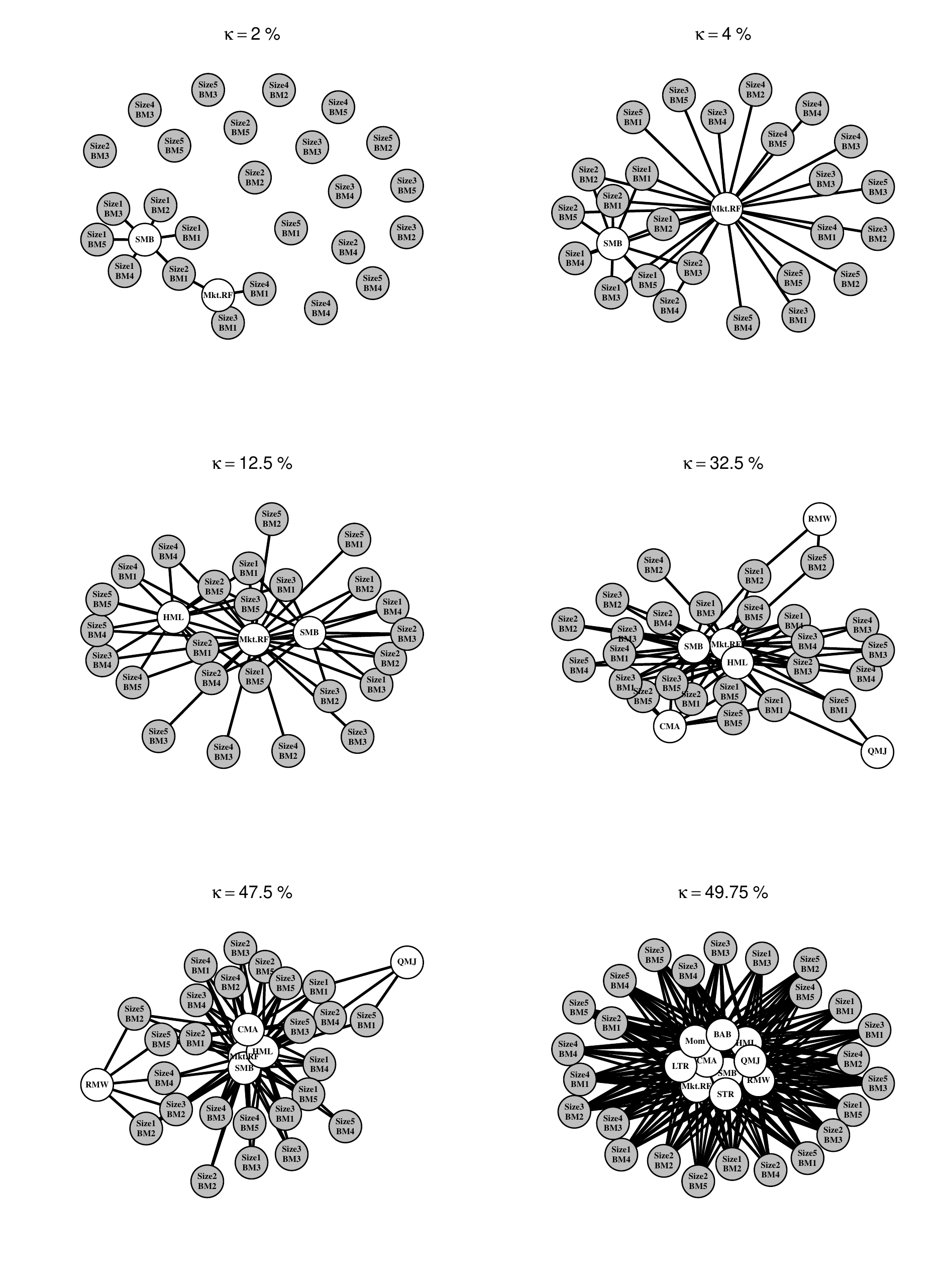}}
  \caption{Sequence of selected models for varying threshold level $\kappa$ under the assumption of \textbf{random predictors}.}
  \label{graphseq25}  
\end{figure}

 In figure \ref{graphseq25}, we examine how different choices of the $\kappa$ threshold change the selected set of risk factors.  In this analysis, there is a tradeoff between the posterior probability of being ``close" to the saturated model and the utility's preference for sparsity.  When the threshold is low ($\kappa=2,4,$ and $12.5$\%) the summarization procedure selects relatively sparse graphs with up to three factors (Mkt.RF, HML, and SMB).  The market (Mkt.RF) and size (SMB) factors appear first, connected to a small number of the test assets ($\kappa=2$\%).  As the threshold is increased, the point summary becomes denser and correspondingly more predictively accurate (as measured by the utility function).  The value factor (HML) enters at $\kappa=12.5$\% and quality minus junk (QMJ), investment (CMA), and profitability (RMW) factors enter at $\kappa=32.5$\%.  The graph for $\kappa=32.5$\% excluding QMJ is essentially the new five factor model proposed by \cite{FF5}.  The five Fama-French factors (plus OMJ with three connections) persist up to the $\kappa=47.5\%$ threshold.  This indicates that, up to a high posterior probability, the five factor model of \cite{FF5} does no worse than an asset pricing model with all ten factors connected to all test assets. 
 
 Notice also that our summarization procedure displays the specific relationship between the factors and test assets through the connections.  Using this approach, the analyst is able to identify which predictors drive variation in which responses and at what thresholds they may be relevant.  This feature is significant for summarization problems where individual characteristics of the test portfolios and their joint dependence on the risk factors is may be a priori unclear.
 
 As $\kappa$ approaches the $50\%$ threshold ($\kappa=49.75\%$ in figure \ref{graphseq25}), the model summary includes all ten factors. Requesting a summary with this level of certainty results in little sparsification.  However, compared to the nearby $\kappa=47.5\%$ model with only six factors, we also now know that the remaining four contribute little to our utility.  These factors are betting against beta (BAB), momentum (Mom), long term reversal (LTR), and short term reversal (STR).  Sparse posterior summarization applied in this context allows an analyst to study the impact of risk factors on pricing while taking uncertainty into account. Coming to a similar conclusion via common alternative techniques (e.g., component-wise ordinary least squares combined with thresholding by $t$-statistics) is comparatively ad hoc; our method is simply a perspicuous summary of a posterior distribution. Likewise, applying sparse regression techniques based on $\ell_1$ penalized likelihood methods would not take into account the residual correlation $\Psi$, nor would that approach naturally accommodate random predictors.

%\begin{figure}[H]
%  \includegraphics[scale=.65]{graph-25PortGenPsi}
%  \caption{The selected model for 25 size/value portfolios modeled by the 10 factors.  The responses and predictors are colored in gray and white, respectively.  Edges represent nonzero components of the optimal action, $\boldsymbol{\gamma}$.}
%  \label{graph25}
%\end{figure}

\subsubsection{Fixed predictors}

In this section, we consider posterior summarization with the loss function derived under the assumption of \textit{fixed predictors}.  The analogous loss function when the predictor matrix is fixed is:
\begin{equation}\label{Lossfix}	
\begin{split}
%\underline{\text{Fixed } \textbf{X}\text{:}}
\mathcal{L}_{\lambda}(\boldsymbol{\gamma}) &=  \norm{ \left[Q_{f}^{T} \otimes L^{T}\right]\text{\bf vec}(\boldsymbol{\gamma}) - \text{\bf vec}(L^{-1}A_{f}Q_{f}^{-T})  }_{2}^{2} + \lambda\norm{ \text{\bf vec}(\boldsymbol{\gamma})}_{1},\\
%\mathcal{L}_{\lambda}(\boldsymbol{\gamma}) =  \norm{ \left[Q^{T} \otimes L^{T}\right]\text{\bf vec}(\boldsymbol{\gamma}) - \text{\bf vec}(L^{-1}AQ^{-T})  }_{2}^{2} + \lambda\norm{ \text{\bf vec}(\boldsymbol{\gamma})}_{1}	
 \end{split}
\end{equation}
with $Q_{f}Q_{f}^{T} = \textbf{X}^{T}\textbf{X}$, $A_{f}=\mathbb{E}[\Omega\tilde{\textbf{Y}}^{T}\textbf{X}]$, and $M=\overline{\Omega}=LL^{T}$; compare to (\ref{moments}) and (\ref{lasso_form}). The derivation of (\ref{Lossfix}) is similar to the presentation in Section \ref{DSS} and may be found in the Appendix.  
The corresponding version of the loss gap is
\begin{equation}
\Delta_{\lambda} = \mathcal{L}(\tilde{\textbf{Y}},\textbf{X},\Theta,\boldsymbol{\gamma}_{\lambda}^*) - \mathcal{L}(\tilde{\textbf{Y}},\textbf{X},\Theta,\boldsymbol{\gamma}^*). 
\end{equation}
which has distribution induced by the posterior over $(\tilde{\textbf{Y}}, \Theta)$ rather than $(\tilde{Y}, \tilde{X}, \Theta)$ as before. By fixing $\textbf{X}$, the posterior of $\Delta_{\lambda}$ has smaller dispersion
%Uncertainty in loss \ref{Lossfix} is driven by the posterior and unknown future response realizations $\tilde{\textbf{Y}}$ through $A_{f}$ and $L$.  In contrast, loss \ref{lasso_form} includes uncertainty via the posterior and future realizations of \textit{both} $\tilde{Y}$ and $\tilde{X}$ through $A$, $Q$, and $L$.  Relative to the random predictor loss, this fixed predictor version will 
which results in denser summaries for the same level of $\kappa$. For example, compare how dense Figure \ref{graph25XFIX} is relative to Figure \ref{graph25}. The denser graph in Figure \ref{graph25XFIX} contains nine out of ten potential risk factors compared to just three in Figure \ref{graph25}, which correspond to the Fama-French factors described in \cite{FF3}. Recall, both graphs represent the sparsest model such that the probability of being no worse than the saturated model is greater than $\kappa = 12.5\%$ --- the difference is that one of the graphs defines ``worse-than'' in terms of a fixed set of risk factor returns while the other acknowledge that those returns are themselves uncertain in future periods.

% In other words, accounting for predictor uncertainty is important for avoiding ``over-fitting" as demonstrated in the densely selected graph \ref{graph25XFIX}. This difference in uncertainty characterization is most clearly seen in figure \ref{Lossgraph25XFIX} where the 75\% posterior intervals are much smaller than its ``random predictors" counterpart in figure \ref{Lossgraph25}.

%The solution path and selected graph are shown in figures \ref{Lossgraph25XFIX} and \ref{graph25XFIX}, respectively.  At the 75\% uncertainty level, the corresponding graph for comparison (where covariate uncertainty is taken into account) is graph \ref{graph25}.  The differences between these graphs is striking.  The graph selected under the fixed covariate assumption has many more connections between test assets and factors.  As can be seen in figure \ref{Lossgraph25XFIX}, the selected model is in the right tail of the solution path.  Additionally, nine out of ten potential risk factors appear in the graph, giving the analyst little indication of which \textit{small} subset of factors may be relevant for asset pricing.  We can understand these differences by recalling the two loss functions used for selection.

\begin{figure}[H]
\centerline{
\includegraphics[scale=.8]{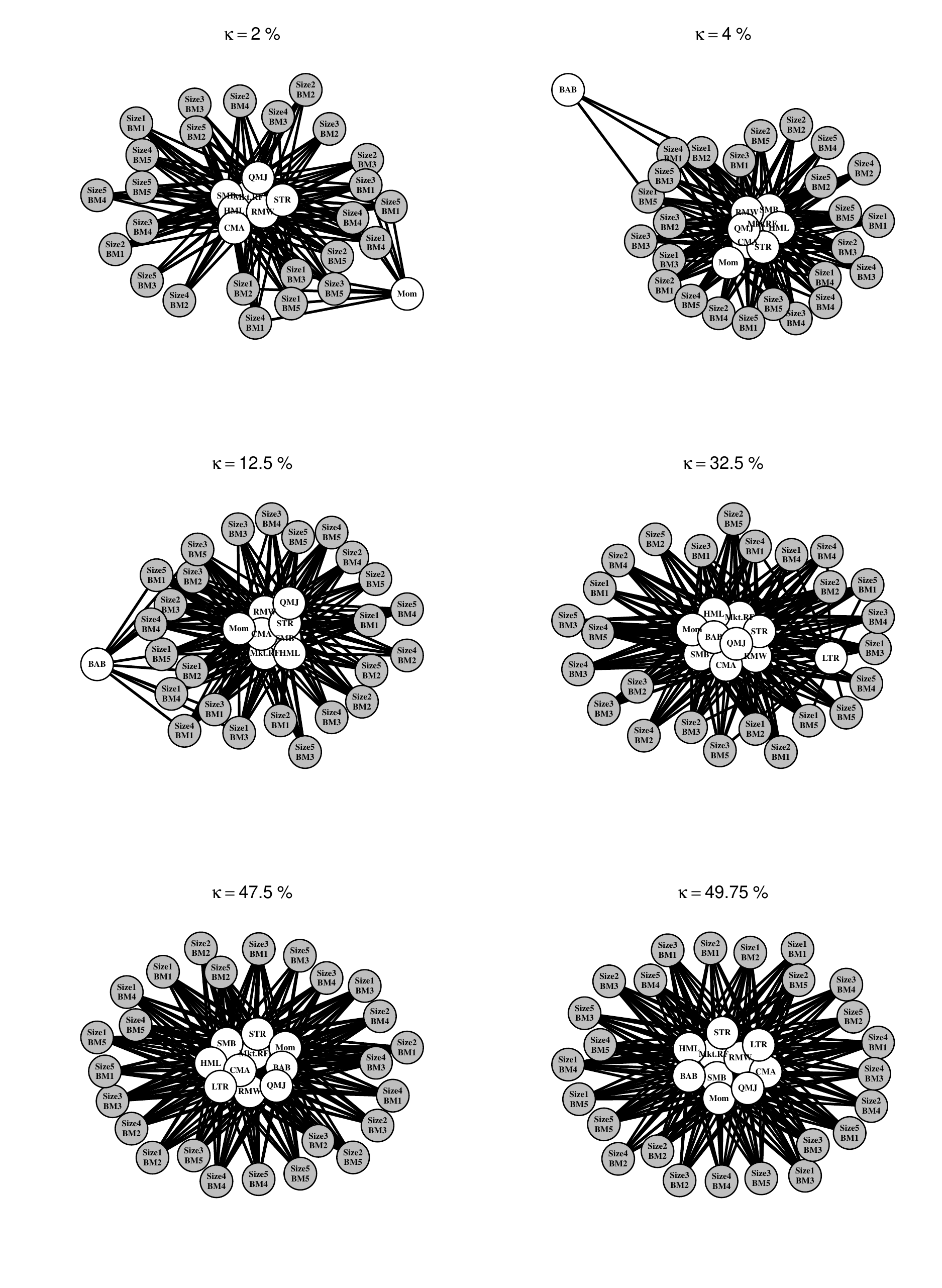}}
\vspace{-12mm}
  \caption{Sequence of selected models for varying threshold level $\kappa$ under the assumption of \textbf{fixed predictors}.}
  \label{graphseq25XFIX}  
\end{figure}

Figure \ref{graphseq25XFIX} demonstrates this problem for several choices of the uncertainty level.  Regardless of the uncertainty level chosen, the selected models contain most of the ten factors and many edges.  In fact, it is difficult to distinguish even the $\kappa=2\%$ and $\kappa=49.75\%$ models.

\begin{figure}[H]
\centering
  \includegraphics[scale=.47]{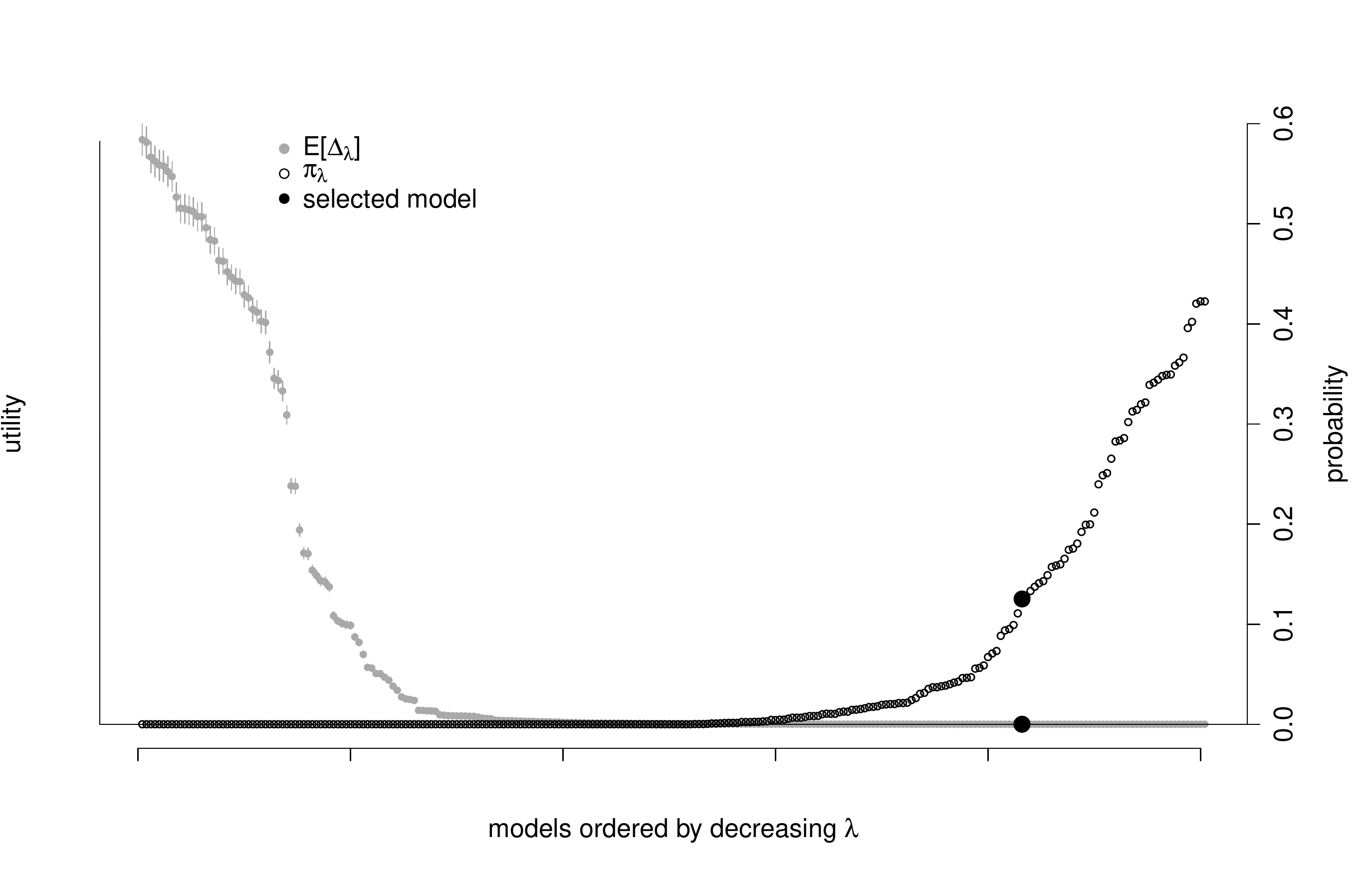}
  \caption{Evaluation of $\Delta_{\lambda}$ and $\pi_{\lambda}$ along the solution path for the 25 size/value portfolios modeled by the 10 factors assuming \textbf{fixed predictors (factors)}. An analyst may use this plot to select a particular model. Uncertainty bands are 75\% posterior intervals on the $\Delta_{\lambda}$ metric. The large black dot represents the model selected in Figure \ref{graph25XFIX}.}
  \label{Lossgraph25XFIX}
  \includegraphics[scale=.65]{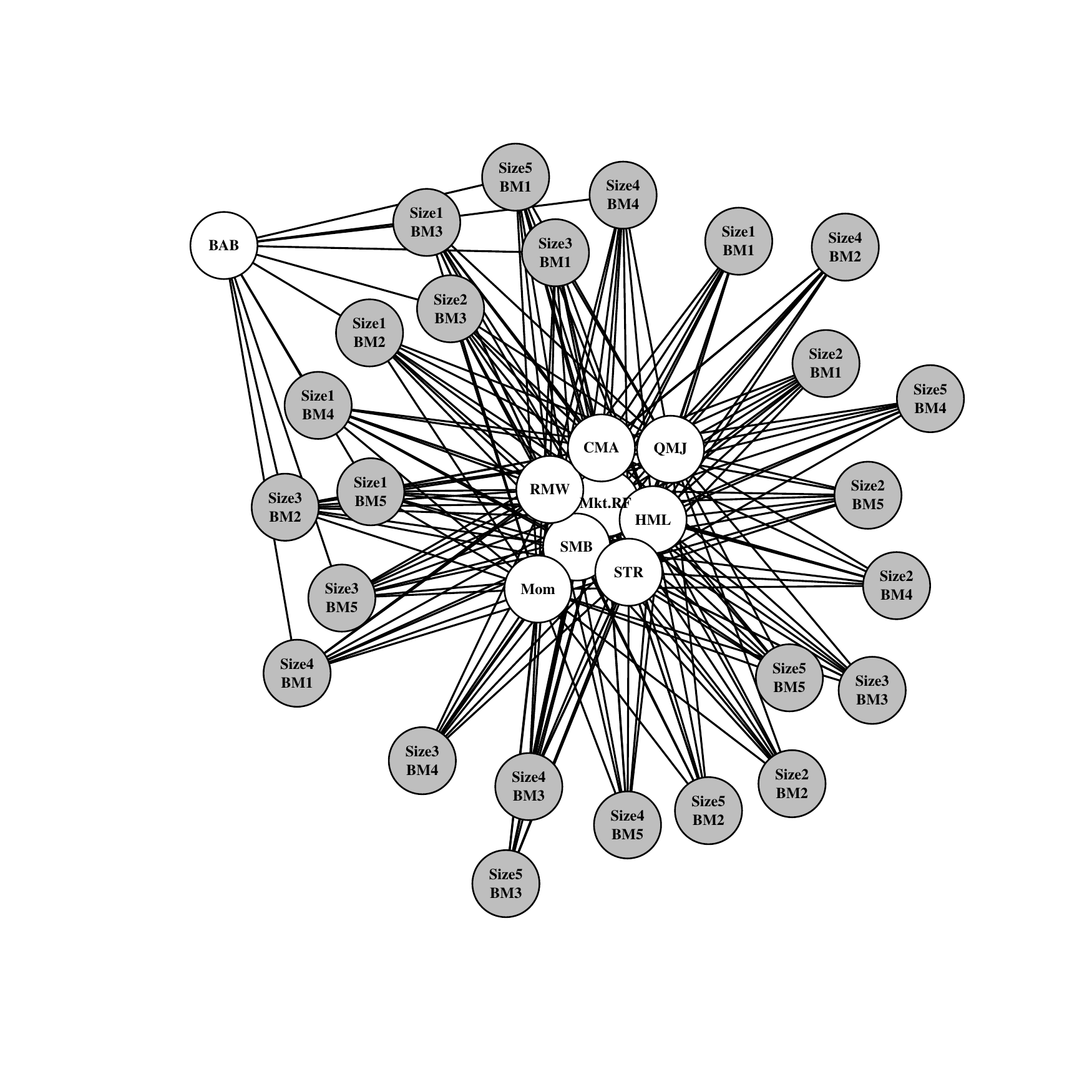}\label{graph25XFIX}
  \vspace{-10mm}
  \caption{The selected model for 25 size/value portfolios modeled by the 10 factors when \textbf{uncertainty in future factor returns is not taken into account}.  The responses and predictors are colored in gray and white, respectively.  Edges represent nonzero components of the optimal action, $\boldsymbol{\gamma}$.}\label{graph25XFIX}
\end{figure}

\subsubsection{Alternative prior analysis}

\begin{figure}[H]
\centering
  \includegraphics[scale=.47]{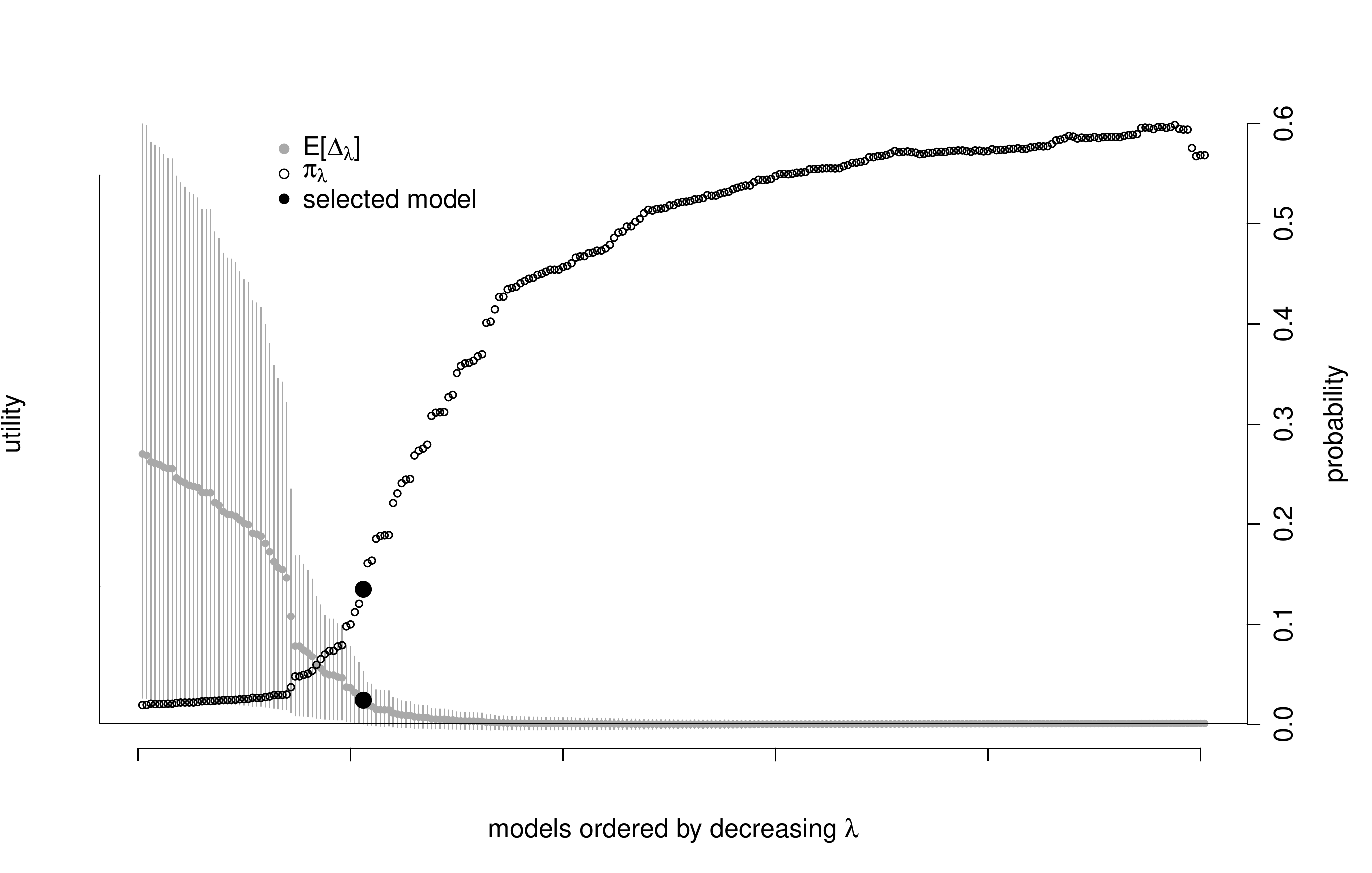}
  \caption{Evaluation of $\Delta_{\lambda}$ and $\pi_{\lambda}$ along the solution path for the 25 size/value portfolios modeled by the 10 factors with alternative prior. An analyst may use this plot to select a particular model. Uncertainty bands are 75\% posterior intervals on the $\Delta_{\lambda}$ metric. The large black dot represents the model selected in Figure \ref{graph25MF}.}
  \label{Lossgraph25MF}
%\end{figure}
%
%\begin{figure}[H]
%\centering
  \includegraphics[scale=.65]{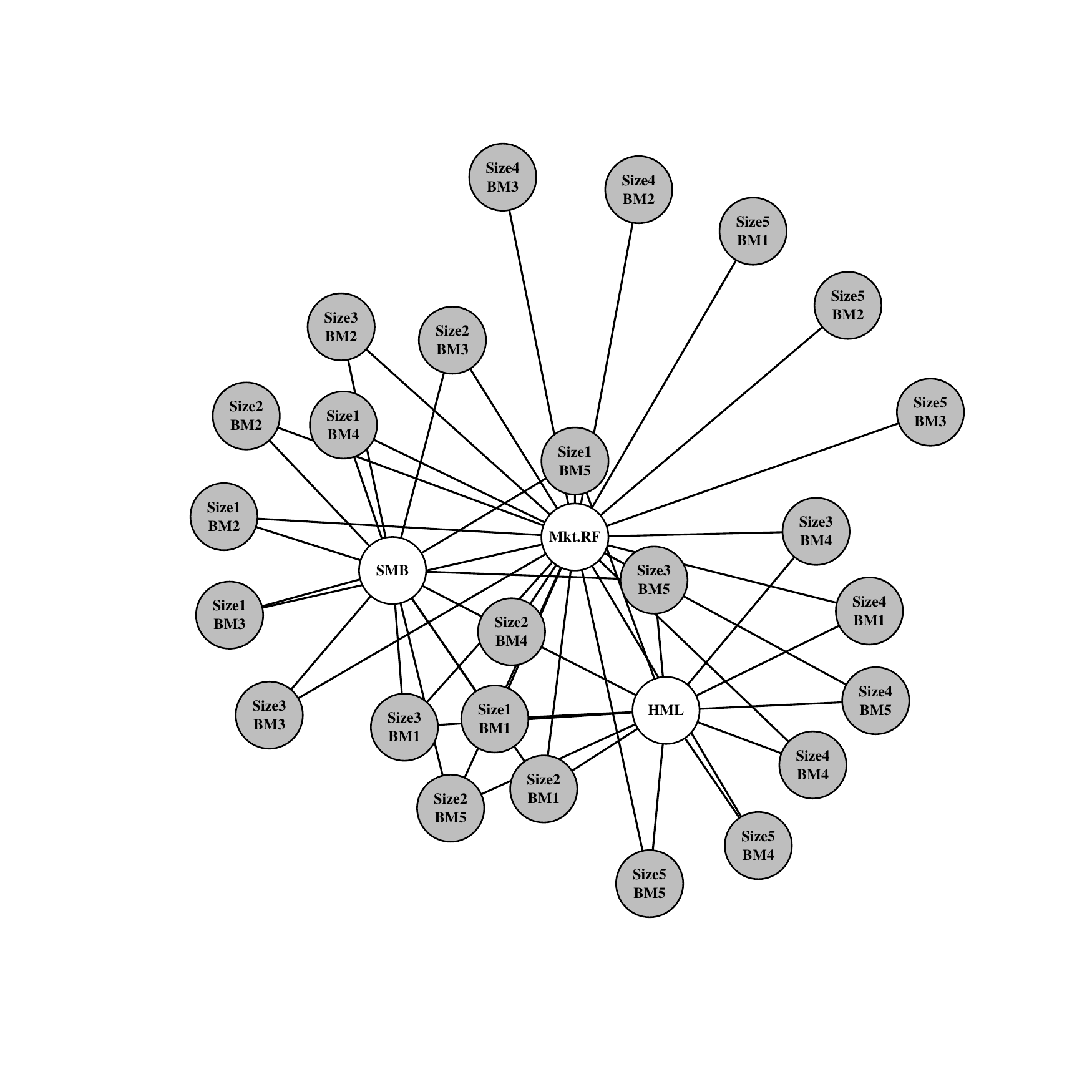}
  \vspace{-20mm}
  \caption{The selected model for 25 size/value portfolios modeled by the 10 factors with alternative prior.  The responses and predictors are colored in gray and white, respectively.  Edges represent nonzero components of the optimal action, $\boldsymbol{\gamma}$.}
  \label{graph25MF}
\end{figure}

Here, we consider how our posterior summaries change as a function of using a different posterior, based on a different choice of prior. Specifically, in this section we do not employ model selection point-mass priors on the elements of $\boldsymbol{\beta}$ as we did in the above analysis. These results are displayed in figures \ref{Lossgraph25MF} and \ref{graph25MF}. Broadly, the same risk factors are flagged as important ---
%As a final exercise, we consider a modified modeling procedure for characterizing uncertainty that demonstrates the flexibility of our methodology by decoupling the posterior inference in {\it step 1} from the summarization in {\it steps 2} and {\it 3}. Any change to the definitions of $p(Y|X)$, $p(X)$ and $p(\Theta)$ can be incorporated to our strategy as long as posterior draws from $p(\Theta | Y, X)$ are available. In figures \ref{Lossgraph25MF}, \ref{graph25MF}, \ref{Lossgraph25MFXFIX}, and \ref{graph25MFXFIX} we fix $\alpha_{i}=1 \hspace{2mm} \forall i$ in our matrix-variate MCMC ($\alpha$ is the binary vector specifying a particular linear model, $M_{\alpha}$). This amounts to independently sampling the regression parameters of each test asset on \textbf{all} 10 factors.  We call this \textit{weak shrinkage} since factor selection is not influenced through sampling of $\alpha$. 
 the market factor followed by the size (SMB) and value (HML) factors.  One notable difference is that the quality minus junk (QMJ), investment (CMA), and profitability (RMW) factors appear at smaller levels of $\kappa$. This result is intuitive in the sense that point-mass priors demand stronger evidence for a variable to impact the posterior means defining the loss function. Without the strong shrinkage imposed by the point-mass priors, these risk factors show up more strongly in the posterior and hence in the posterior summary. In each case, the three Fama and French factors from \cite{FF3} predictably appear and seem to be the only relevant factors for pricing these 25 portfolios.
 
\begin{figure}[H]
\centering
  \includegraphics[scale=.47]{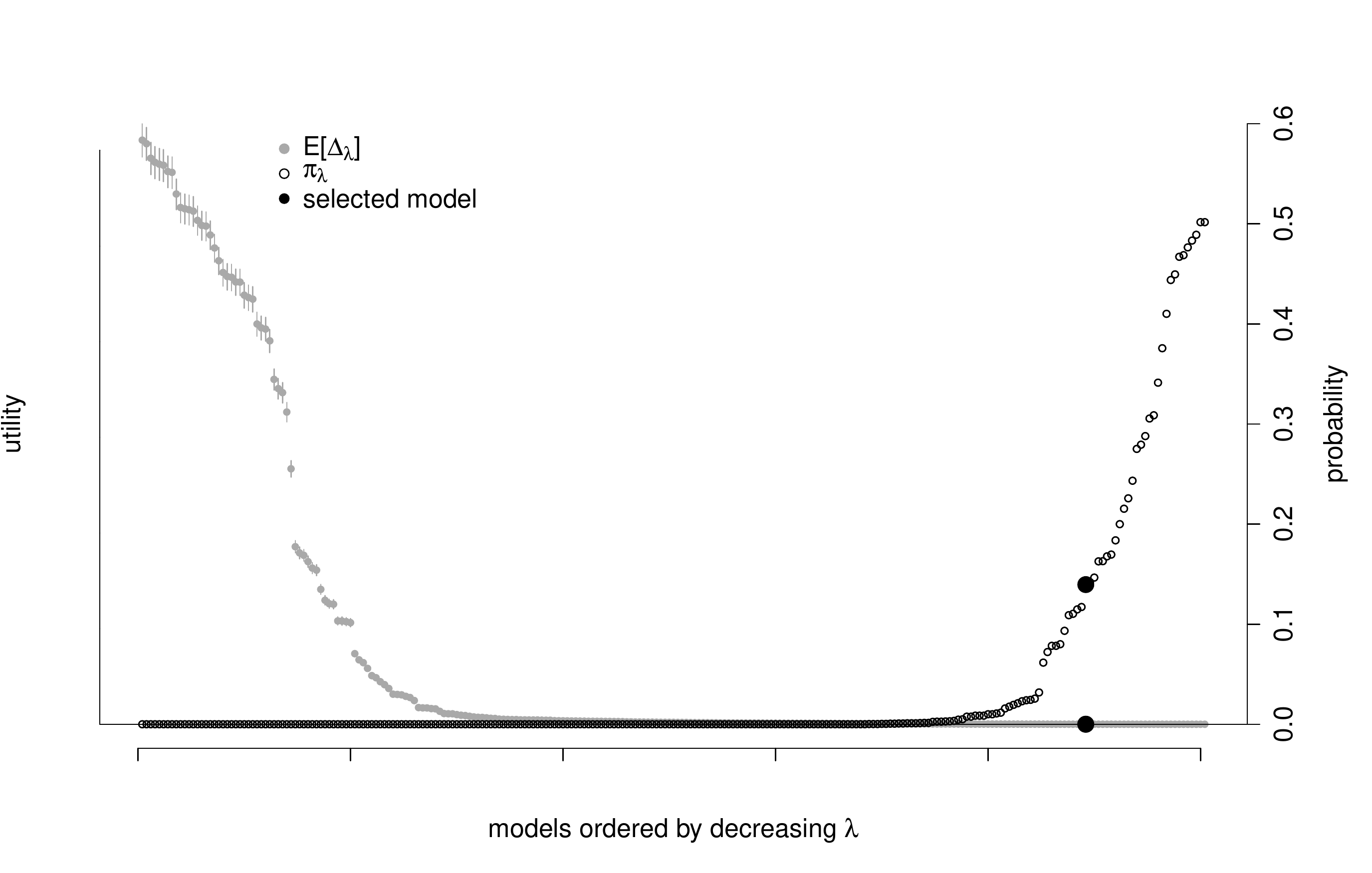}
  \caption{Evaluation of $\Delta_{\lambda}$ and $\pi_{\lambda}$ along the solution path for the 25 size/value portfolios modeled by the 10 factors assuming \textbf{fixed predictors (factors)} and with alternative prior. An analyst may use this plot to select a particular model. Uncertainty bands are 75\% posterior intervals on the $\Delta_{\lambda}$ metric. The large black dot represents the model selected in Figure \ref{graph25MFXFIX}.}
  \label{Lossgraph25MFXFIX}
%\end{figure}
%
%\begin{figure}[H]
%\centering
  \includegraphics[scale=.65]{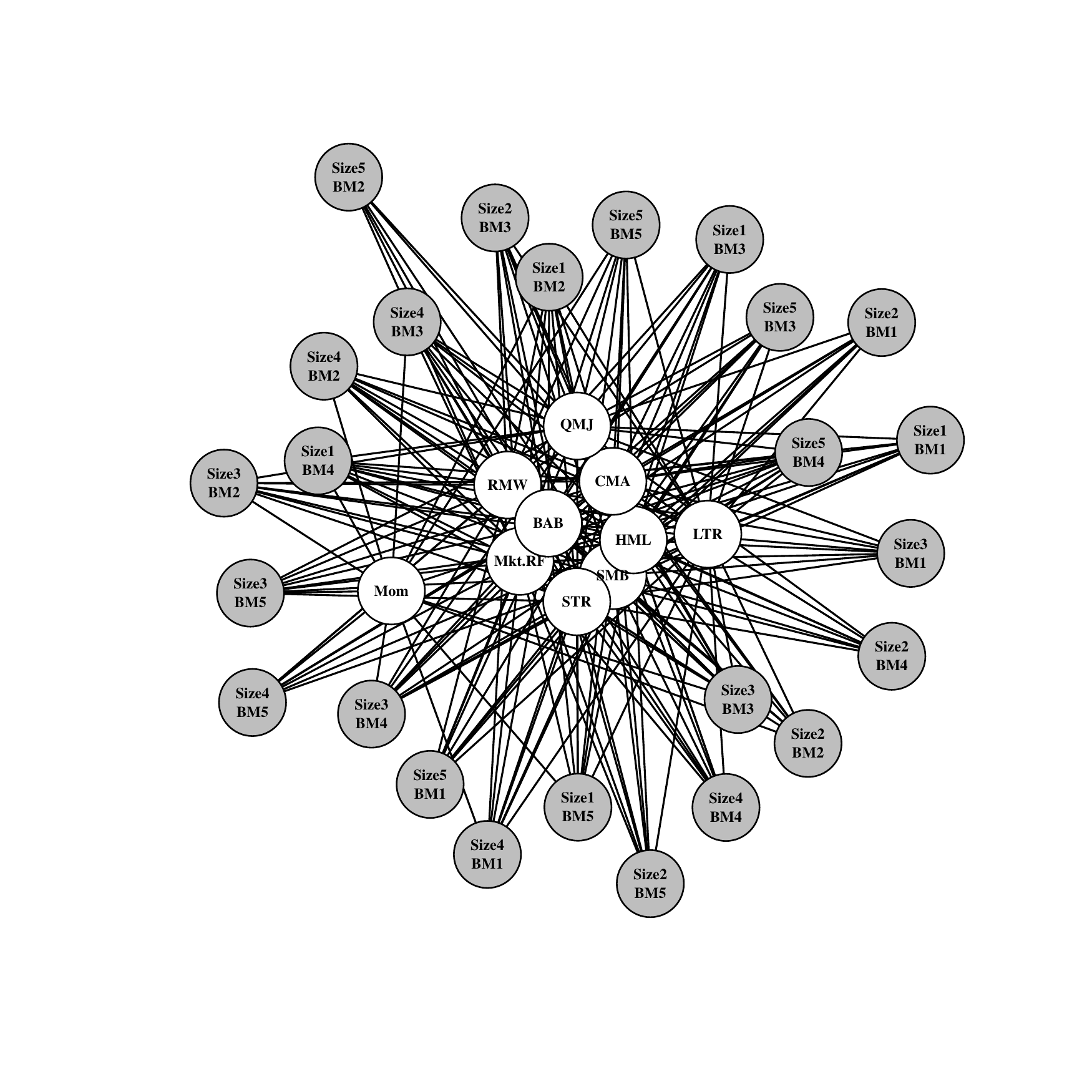}
  \vspace{-10mm}
  \caption{The selected model for 25 size/value portfolios modeled by the 10 factors when \textbf{uncertainty in future factor returns is not taken into account} and with alternative prior.  The responses and predictors are colored in gray and white, respectively.  Edges represent nonzero components of the optimal action, $\boldsymbol{\gamma}$.}
  \label{graph25MFXFIX}
\end{figure}

Similarly, the weaker shrinkage model in the fixed predictor version (figures \ref{Lossgraph25MFXFIX} and \ref{graph25MFXFIX}) yields yet denser summaries (for a given level of $\kappa$).

%We hope that this example demonstrates how our proposed methodology may be used to identify factors for asset pricing.  

\subsubsection{Comparison of four scenarios at fixed $\kappa$}

\begin{figure}[H]
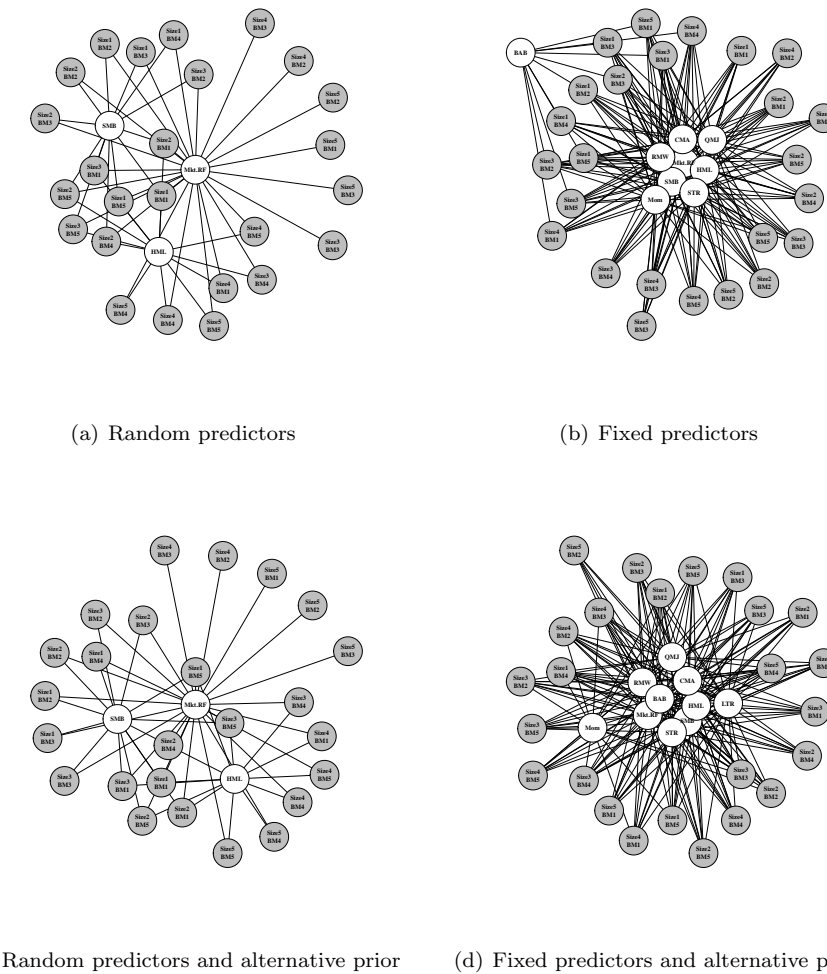

\centering
	 \subfigure[Random predictors]{\includegraphics[scale=.35]{graph-25PortGenPsi}}
	 \subfigure[Fixed predictors]{\includegraphics[scale=.35]{graph-25PortXFIXGenPsi}}
	 \subfigure[Random predictors and alternative prior]{\includegraphics[scale=.35]{graph-25PortMODELFIXEDGenPsi}}
	 \subfigure[Fixed predictors and alternative prior]{\includegraphics[scale=.35]{graph-25PortMODELFIXEDXFIXGenPsi}}
	 \caption{Comparison of selected models under four scenarios.}\label{summarygraph}
\end{figure}

The selected summary graphs for the four scenarios are displayed together for comparison in figure \ref{summarygraph}.  Observe that graphs (c) and (d) selected under the alternative prior are marginally denser than their counterparts (a) and (b) under the point-mass model selection prior.  However, the assumption of random predictors results in notably sparser summaries -- graphs (a) and (c) are much sparser than (b) and (d). These comparisons emphasize the impact that incorporating random predictors may have on a variable selection procedure; especially the present approach where we extract point summaries from a posterior by utilizing uncertainty in all unknowns $(\tilde{Y},\tilde{X},\Theta)$.  

\section{Conclusion}

In this paper, we propose a general model selection procedure for multivariate linear models when future realizations of the predictors are unknown.  Such models are widely used in many areas of science and economics, including genetics and asset pricing. Our utility-based sparse posterior summary procedure is a multivariate extension of the ``decoupling shrinkage and selection" methodology of \cite{HahnCarvalho}. The approach we develop has three steps: (\textit{i}) fit a Bayesian model, (\textit{ii}) specify a utility function with a sparsity-inducing penalty term and optimize its expectation, and (\textit{iii}) graphically summarize the posterior impact (in terms of utility) of the sparsity penalty.  Our utility function is based on the kernel of the conditional distribution responses given the predictors and can be formulated as a tractable convex program. We demonstrate how our procedure may be used in asset pricing under a variety of modeling choices.

The remainder of this discussion takes a step back from the specifics of the seemingly unrelated regressions model and considers a broader role for utility-based posterior summaries.

A paradox of applied Bayesian analysis is that posterior distributions based on relatively intuitive models like the SUR model are often just as complicated as the data itself. For Bayesian analysis to become a routine tool for practical inquiry, methods for summarizing posterior distributions must be developed apace with the models themselves. A natural starting point for developing such methods is decision theory, which suggests developing loss functions specifically geared towards practical posterior summary. As a matter of practical data analysis, articulating an apt loss function has been sorely neglected relative to the effort typically lavished on the model specification stage, specifically prior specification. Ironically (but not surprisingly) our application demonstrates that one's utility function has a dominant effect on the posterior summaries obtained relative to which prior distribution is used.

This paper makes two contributions to this area of ``utility design''. First, we propose that the likelihood function has a role to play in posterior summary apart from its role in inference. That is, one of the great practical virtues of likelihood-based statistics is that the likelihood serves to summarize the data by way of the corresponding point estimates. By using the log-density as our utility function applied to {\em future} data, we revive the fundamental summarizing role of the likelihood. Additionally, note that this approach allows three distinct roles for parameters. First, all parameters of the model appear in defining the posterior predictive distribution. Second, some parameters appear in {\em defining} the loss function; $\Psi$ plays this role in our analysis. Third, some parameters define the action space. In this framework there are no ``nuisance'' parameters that vanish from the estimator as soon as a marginal posterior is obtained. Once the likelihood-based utility is specified, it is a natural next step to consider augmenting the utility to enforce particular features of the desired point summary. For example, our analysis above was based on a utility that explicitly rewards sparsity of the resulting summary. A traditional instance of this idea is the definition of high posterior density regions, which are defined as the {\em shortest, contiguous} interval that contains a prescribed fraction of the posterior mass.

Our second contribution is to consider not just one, but a range, of utility functions and to examine the posterior distributions of the corresponding posterior loss. Specifically, we compare the utility of a sparsified summary to the utility of the optimal non-sparse summary. Interestingly, these utilities are random variables themselves (defined by the posterior distribution) and examining their distributions provides a fundamentally Bayesian way to measure the extent to which the sparsity preference is driving one's conclusions. The idea of comparing a hypothetical continuum of decision-makers based on the posterior distribution of their respective utilities represents a principled Bayesian approach to exploratory data analysis.  This is an area of ongoing research.

\appendix

\section{Matrix-variate Stochastic Search}

\subsection{Model fitting: The marginal and conditional distributions}

The future values of the response and covariates are unknown.  Acknowledging this uncertainty is important in the overall decision of which covariates to select and is a necessary ingredient of the selection procedure.  As our examples consider financial asset return data, we choose to model the marginal distribution of the covariates via a latent factor model detailed in \cite{murray2013bayesian}. The responses are modeled conditionally on the covariates via a matrix-variate stochastic search which is a multivariate of extension of stochastic search variable selection (SSVS) from \cite{GeorgeandMcCulloch}.  Recalling the block structure for the covariance of the full joint distribution of $(X,Y)$:

\begin{align}
\Sigma
= 
\left[
\begin{array}{c|c}
\boldsymbol{\beta}^{T}\Sigma_{x}\boldsymbol{\beta} + \Psi & (\Sigma_{x}\boldsymbol{\beta})^{T} \\
\hline
\Sigma_{x}\boldsymbol{\beta}
 & 
\Sigma_{x} \\
\end{array}
\right],
\end{align}we obtain posterior samples of $\Sigma$ by sampling the conditional model parameters using a matrix-variate stochastic search algorithm (described below) and sampling the covariance of $X$ from a latent factor model where it is marginally normally distributed.  To reiterate our procedure is

\begin{itemize}
	\item $\Sigma_{x}$ is sampled from independent latent factor model,
	\item $\boldsymbol{\beta}$ is sampled from matrix-variate MCMC,
	\item $\Psi$ is sampled from matrix-variate MCMC.
\end{itemize}

\subsubsection{Modeling a full residual covariance matrix}

In order to sample a full residual covariance matrix, we augment the predictor matrix with a latent factor $f$ by substituting $\epsilon_{j} = b_{j}f + \tilde{\epsilon}_{j}$:

\begin{equation}\label{modelfirstA}
\begin{split}
	Y_{j} &= \beta_{j1}X_{1} + \cdots + \beta_{jp}X_{p} + b_{j}f + \tilde{\epsilon}_{j}, \;\;\;\;\; \tilde{\boldsymbol{\epsilon}} \sim \mbox{N}(0, \tilde{\Psi}),
\end{split}
\end{equation}where $\tilde{\Psi}$ is now diagonal. Assuming that $f \sim N(0,1)$ is shared among all response variables $j$ and $\textbf{b} \in \mathbb{R}^{qx1}$ is a vector of all coefficients $b_{j}$, the total residual variance may be expressed as:

\begin{equation}
	\begin{split}
		\Psi = \textbf{b}\textbf{b}^{T} + \tilde{\Psi}.
	\end{split}
\end{equation}We incorporate this latent factor model into the matrix-variate MCMC via a simple Gibbs step to draw posterior samples of $f$.  This augmentation allows us to draw samples of $\Psi$ that are not constrained to be diagonal.

\subsubsection{Modeling the marginal distribution: A latent factor model}

We model covariates via a latent factor model of the form:

\begin{equation}
	\begin{split}
		\textbf{X}_{t} &= \mu_{x} + \textbf{B}\textbf{f}_{t} + \textbf{v}_{t}
		\\
		\textbf{v}_{t} \sim \text{N}(0,\mathbf{\Lambda}), \hspace{4mm} &\textbf{f}_{t} \sim \text{N}(0,\mathbb{I}_{k}), \hspace{4mm} \mu_{x} \sim \text{N}(0,\Phi) 
	\end{split}
\end{equation}where $\Lambda$ is assumed diagonal and the set of $k$ latent factors $f_{t}$ are independent.  The covariance of the covariates is constrained by the factor decomposition and takes the form:

\begin{equation}
	\begin{split}
		\Sigma_{x} = \textbf{B}\textbf{B}^{T} + \Lambda.
	\end{split}
\end{equation} 
Recall that this is only a potential choice for the $p(X)$ and it is chosen here primarily motivated by the applied context where financial assets tend to depend across each other through common factors. Our variable selection procedure would follow if any other choice was made at this point. 
To estimate this model, a convenient, efficient choice is the R package {\tt bfa} \citep{bfa}.  The software allows us to sample the marginal covariance as well as the marginal mean via a simple Gibbs step assuming a normal prior on $\mu_{x}$.

\subsubsection{Modeling the conditional distribution: A matrix-variate stochastic search}

We model the conditional distribution, $Y \vert X$, by developing a multivariate extension of stochastic search variable selection of \cite{GeorgeandMcCulloch}.  Recall that the conditional model is: $\textbf{Y} - \textbf{X} \boldsymbol \beta \sim \mathcal{N}\left(\mathbb{I}_{N \times N}, \hspace{1mm} \Psi_{q \times q} \right)$.  In order to sample different subsets of covariates (different models) during the posterior simulations, we introduce an additional parameter $\alpha \in \mathbb{R}^{p}$ that is a binary vector identifying a particular model. In other words, all entries $i$ for which $\alpha_{i}=1$ denote covariate $i$ as included in model $M_{\alpha}$.  Specifically, we write the model identified by $\alpha$ as $M_{\alpha}: \textbf{Y} - \textbf{X}_{\alpha} \boldsymbol{\beta}_{\alpha} \sim \mathcal{N}\left(\mathbb{I}_{N \times N}, \hspace{1mm} \Psi_{q \times q} \right)$.  As in \cite{GeorgeandMcCulloch}, we aim to explore the posterior on the model space, $\textbf{P}\left(M_{\alpha} \hspace{1mm} \vert \hspace{1mm} \textbf{Y} \right)$.  Our algorithm explores this model space by calculating a Bayes factor for a particular model $M_{\alpha}$.  Given that the response $\textbf{Y}$ is matrix instead of a vector, we derive the Bayes factor as a product of vector response Bayes factors.  This is done by separating the marginal likelihood of the response matrix as a product of marginal likelihoods across the separate vector responses.  This derivation requires our priors to be independent across the responses and is shown in the Appendix.  It is important to note that we do not run a standard SSVS on each univariate response regression separately. Instead, we generalize \cite{GeorgeandMcCulloch} and require all covariates to be included or excluded from a model for each of the responses \textit{simultaneously}. 

The marginal likelihood requires priors for the parameters $\boldsymbol{\beta}$ and $\sigma$ parameters in our model.  We choose the standard g-prior for linear models because it permits an analytical solution for the marginal likelihood integral \citep{Z1,Z3,Liangetal08}.  
%These priors come from the work of Zellner, Jeffreys, and Siow, and allow for a non-informative prior for $\sigma$ and conditional g-prior on $\boldsymbol \beta_{\gamma}^{i}$ \cite{Z1}, \cite{Z2}, \cite{Z3}, \cite{J1}.

Our Gibbs sampling algorithm directly follows the stochastic search variable selection procedure described in \cite{GeorgeandMcCulloch} using these calculated Bayes factors, now adapted to a multivariate setting.  The aim is to scan through all possible covariates and determine which ones to include in the model identified through the binary vector $\alpha$.  At each substep of the MCMC, we consider an individual covariate $i$ within a specific model and compute its inclusion probability as a function of the model's prior probability and the Bayes factors:

\begin{equation*}
\begin{split}
	p_{i} = \frac{B_{a0} \textbf{P}\left(M_{\alpha_{a}}\right)}{B_{a0} \textbf{P}\left(M_{\alpha_{a}}\right) + B_{b0} \textbf{P}\left(M_{\alpha_{b}}\right)}.
\end{split}	
\end{equation*}The Bayes factor $B_{a0}$ is a ratio of marginal likelihoods for the model with covariate $i$ included and the null model, and $B_{b0}$ is the analogous Bayes factor for the model without covariate $i$. The prior on the model space, $\textbf{P}\left(M_{\gamma}\right)$, can either be chosen to adjust for multiplicity or to be uniform - our results appear robust to both specifications.  In this setting, adjusting for multiplicity amounts to putting equal prior mass on different sizes of models.  In contrast, the uniform prior for models involving $p$ covariates puts higher probability mass on larger models, reaching a maximum for models with ${ p \choose 2 }$ covariates included.  The details of the priors on the model space and parameters, including an empirical Bayes choice of the g-prior hyperparameter, are discussed in the Appendix.

\subsection{Details}

Assume we have observed $N$ realizations of data $(\mathbf{Y},\mathbf{X})$. For model comparison, we calculate the Bayes factor with respect to the null model without any covariates.  First, we calculate a marginal likelihood.  This likelihood is obtained by integrating the full model over $\boldsymbol \beta_{\alpha}$ and $\sigma$ multiplied by a prior, $\pi_{\alpha}\left(\boldsymbol \beta_{\alpha}, \sigma\right) $, for these parameters.  A Bayes factor of a given model $\alpha$ versus the null model, $B_{\alpha 0} = \frac{m_{\alpha}\left(\textbf{R}\right)}{m_{0}\left(\textbf{R}\right)}$ with:

\begin{align} \label{marginal}
m_{\alpha}\left(\mathbf{Y}\right) = \int \textrm{Matrix Normal}_{N,q}\left( \mathbf{Y} \hspace{1mm} \vert \hspace{1mm} \textbf{X}_{\alpha} \boldsymbol \beta_{\alpha}, \hspace{1mm} \mathbb{I}_{N x N}, \hspace{1mm} \tilde{\Psi}_{q \times q} \right) \pi_{\alpha}\left(\boldsymbol \beta_{\alpha}, \sigma_{i}\right) d\boldsymbol \beta_{\alpha} d\sigma_{i}.
\end{align}We assume independence of the priors across columns of $\mathbf{Y}$ so we can write the integrand in (\ref{marginal}) as a product across each individual response vector:

\begin{align*}
m_{\alpha}\left(\mathbf{Y}\right) &= \int \Pi_{i=1}^{q} \hspace{1mm} N_{N}\left( \mathbf{Y}^{i} \hspace{1mm} \vert \hspace{1mm} \textbf{X}_{\alpha} \boldsymbol \beta_{\alpha}^{i}, \hspace{1mm} \sigma_{i}^{2}\mathbb{I}_{N x N}\right) \pi_{\alpha}^{i} \left(\boldsymbol \beta_{\alpha}^{i}, \sigma_{i}\right) d\boldsymbol \beta_{\alpha}^{i} d\sigma_{i}
\\
&\iff
\\
m_{\alpha}\left(\mathbf{Y}\right) &= \int \hspace{1mm} N_{N}\left( \mathbf{Y}^{1} \hspace{1mm} \vert \hspace{1mm} \textbf{X}_{\alpha} \boldsymbol \beta_{\alpha}^{1}, \hspace{1mm} \sigma_{1}^{2}\mathbb{I}_{N x N}\right) \pi_{\alpha}^{1} \left(\boldsymbol \beta_{\alpha}^{1}, \sigma_{1}\right) d\boldsymbol \beta_{\alpha}^{1} d\sigma_{1} \\ & \times \cdots \times \int N_{N}\left( \mathbf{Y}^{q} \hspace{1mm} \vert \hspace{1mm} \textbf{X}_{\alpha} \boldsymbol \beta_{\alpha}^{q}, \hspace{1mm} \sigma_{q}^{2}\mathbb{I}_{N x N}\right) \pi_{\alpha}^{q} \left(\boldsymbol \beta_{\alpha}^{q}, \sigma_{q}\right) d\boldsymbol \beta_{\alpha}^{q} d\sigma_{q}
\\
&= m_{\alpha}\left(\mathbf{Y}^{1}\right) \times \cdots \times m_{\alpha}\left(\mathbf{Y}^{q}\right) 
\\
&= \Pi_{i=1}^{q} m_{\alpha}\left(\mathbf{Y}^{i}\right),
\end{align*}

with:

\begin{align} \label{A5}
\mathbf{Y}^{i} \sim N_{N}\left(\textbf{X}_{\alpha} \boldsymbol \beta_{\alpha}^{i}, \hspace{1mm} \sigma_{i}^{2}\mathbb{I}_{N x N}\right).
\end{align}Therefore, the Bayes factor for this matrix-variate model is just a product of Bayes factors for the individual multivariate normal models.

\begin{align} \label{A6}
B_{\alpha0} = \widetilde{B}_{\alpha0}^{1} \times \cdots \times  \widetilde{B}_{\alpha0}^{q}
\end{align}

with:

\begin{align} \label{A6}
\widetilde{B}_{\alpha0}^{i} = \frac{m_{\alpha}\left(\mathbf{Y}^{i}\right)}{m_{0}\left(\mathbf{Y}^{i}\right)}.
\end{align}

The simplification of the marginal likelihood calculation is crucial for analytical simplicity and for the resulting SSVS algorithm to rely on techniques already developed for univariate response models.  In order to calculate the integral for each Bayes factor, we need priors on the parameters $\boldsymbol{\beta}_{\alpha}$ and $\sigma$.  Since the priors are independent across the columns of $\mathbf{Y}$, we aim to define $\pi_{\alpha}^{i} \left(\boldsymbol \beta_{\alpha}^{i}, \sigma_{i}\right)$ $\forall i \in \{1,...,q\}$, which we express as the product: $\pi_{\alpha}^{i} \left(\sigma_{i}\right) \pi_{\alpha}^{i} \left(\boldsymbol \beta_{\alpha}^{i} \hspace{1mm} \vert \hspace{1mm} \sigma_{i}\right)$.  Motivated by the work on regression problems of Zellner, Jeffreys, and Siow, we choose a non-informative prior for $\sigma_{i}$ and the popular g-prior for the conditional prior on $\boldsymbol \beta_{\alpha}^{i}$, \citep{Z1}, \citep{Z2}, \citep{Z3}, \citep{J1}: 

\begin{align} \label{A7}
\pi_{\alpha}^{i} \left(\boldsymbol \beta_{\alpha}^{i}, \sigma_{i} \hspace{1mm} \vert \hspace{1mm}  g \right) = \sigma_{i}^{-1} \textrm{N}_{k_{\alpha}}\left(\boldsymbol \beta_{\alpha}^{i} \hspace{1mm} \vert \hspace{1mm} \textbf{0}, g_{\alpha}^{i} \sigma_{i}^2 (\textbf{X}_{\alpha}^{T}(\mathbb{I} - N^{-1}\textbf{1}\textbf{1}^{T})\textbf{X}_{\alpha})^{-1}\right).
\end{align}Under this prior, we have an analytical form for the Bayes factor:

\begin{align} \label{A8}
B_{\alpha0} &= \widetilde{B}_{\alpha0}^{1} \times \cdots \times  \widetilde{B}_{\alpha0}^{q}
\\
&= \Pi_{i=1}^{q} \frac{\left(1 +  g_{\alpha}^{i}\right)^{(N-k_{\alpha}-1)/2}}{\left(1 +  g_{\alpha}^{i}\frac{SSE_{\alpha}^{i}}{SSE_{0}^{i}}\right)^{(N+1)/2}},
\end{align}where $SSE_{\alpha}^{i}$ and $SSE_{0}^{i}$ are the sum of squared errors from the linear regression of column $\mathbf{Y}^{i}$ on covariates $\textbf{X}_{\alpha}$ and $k_{\alpha}$ is the number of covariates in model $M_{\alpha}$.  We allow the hyper parameter $g$ to vary across columns of $\mathbf{Y}$ and depend on the model, denoted by writing, $g_{\alpha}^{i}$.
\\
\\
We aim to explore the posterior of the model space, given our data:
\begin{align} \label{A9}
\textbf{P}\left(M_{\alpha} \hspace{1mm} \vert \hspace{1mm} \mathbf{Y} \right) = \frac{B_{\alpha0} \textbf{P}\left(M_{\alpha}\right)}{\Sigma_{\alpha} B_{\alpha0} \textbf{P}\left(M_{\alpha}\right)},
\end{align}where the denominator is a normalization factor.  In the spirit of traditional stochastic search variable selection \cite{OnSam}, we propose the following Gibbs sampler to sample this posterior.

\subsection{Gibbs Sampling Algorithm}

Once the parameters $\boldsymbol \beta_{\alpha}$ and $\sigma$ are integrated out, we know the form of the full conditional distributions for $\alpha_{i} \hspace{1mm} \vert \hspace{1mm} \alpha_{1}, \cdots, \alpha_{i-1}, \alpha_{i+1}, \cdots, \alpha_{p}$.  We sample from these distributions as follows:

\begin{enumerate}
\item Choose column $\mathbf{Y}^{i}$ and consider two models $\alpha_{a}$ and $\alpha_{b}$ such that:
\begin{align*}
\alpha_{a} = (\alpha_{1}, \cdots, \alpha_{i-1}, 1, \alpha_{i+1}, \cdots, \alpha_{p})
\\
\alpha_{b} = (\alpha_{1}, \cdots, \alpha_{i-1}, 0, \alpha_{i+1}, \cdots, \alpha_{p})
\end{align*}

\item For each model, calculate $B_{a0}$ and $B_{b0}$ as defined by (\ref{A8}).

\item Sample 
\begin{align*}
\alpha_{i} \hspace{1mm} \vert \hspace{1mm} \alpha_{1}, \cdots, \alpha_{i-1}, \alpha_{i+1}, \cdots, \alpha_{p} \sim Ber(p_{i})
\end{align*}

where

\begin{align*}
p_{i} = \frac{B_{a0} \textbf{P}\left(M_{\alpha_{a}}\right)}{B_{a0} \textbf{P}\left(M_{\alpha_{a}}\right) + B_{b0} \textbf{P}\left(M_{\alpha_{b}}\right)},
\end{align*}

\end{enumerate}

Using this algorithm, we visit the most likely models given our set of responses.  Under the model and prior specification, there are closed-form expressions for the posteriors of the model parameters $\beta_{\alpha}$ and $\sigma$.

\subsection{Hyper Parameter for the $g$-prior}

We use a local empirical Bayes to choose the hyper parameter for the $g$-prior in (\ref{A7}).  Since we allow $g$ to be a function of the columns of $\mathbf{Y}$ as well as the model defined by $\alpha$, we calculate a separate $g$ for each univariate Bayes factor in (\ref{A7}) above.  An empirical Bayes estimate of $g$ maximizes the marginal likelihood and is constrained to be non-negative.  From \cite{Liang}, we have:

\begin{align}
\hat{g}_{\alpha}^{EB(i)} &= max\{F_{\alpha}^{i}-1,0\}
\\
F_{\alpha}^{i} &= \frac{R_{\alpha}^{2i} / k_{\alpha}}{(1-R_{\alpha}^{2i}) / (N - 1 - k_{\alpha})}.
\end{align}

For univariate stochastic search, the literature recommends choosing a fixed $g$ as the number of data points \cite{OnSam}.  However, the multivariate nature of our model induced by the vector-valued response makes this approach unreliable. Since each response has distinct statistical characteristics and correlations with the covariates, it is necessary to vary $g$ among different sampled models and responses.  We find that this approach provides sufficiently stable estimation of the inclusion probabilities for the covariates.

\section{Derivation of lasso form}
In this section of the Appendix, we derive the penalized objective (lasso) forms of the utility functions.  After integration over $p(\tilde{Y},\tilde{X}, \Theta \vert \textbf{Y}, \textbf{X})$, the utility takes the form (from equation (\ref{almostlassoform})):

\begin{equation}
	\begin{split}
		\mathcal{L}(\boldsymbol{\gamma}) &= \text{tr}[ M \boldsymbol{\gamma} S \boldsymbol{\gamma}^{T} ] - 2\text{tr}[A\boldsymbol{\gamma}^{T}]  + \lambda \norm{\text{{\bf vec}}(\boldsymbol{\gamma})}_{0},
	\end{split}
\end{equation}where $A=\mathbb{E}[\Omega\tilde{Y}\tilde{X}^{T}]$, $S=\mathbb{E}[\tilde{X}\tilde{X}^{T}] = \overline{\Sigma_{x}}$, and $M=\overline{\Omega}$, and the overlines denote posterior means. Defining the Cholesky decompositions: $M = LL^{T}$ and $S = QQ^{T}$, combining the matrix traces, completing the square with respect to $\boldsymbol{\gamma}$, and converting the trace to the vectorization operator, we obtain:

\begin{equation}
	\begin{split}
		\mathcal{L}(\boldsymbol{\gamma}) &= \text{tr}[M(\boldsymbol{\gamma} S \boldsymbol{\gamma}^{T} - 2M^{-1}A\boldsymbol{\gamma}^{T} ] + \lambda \norm{\text{ {\bf vec}}(\boldsymbol{\gamma})}_{0}
		\\
		&\propto \text{tr}\left[M(\boldsymbol{\gamma} - M^{-1}AS^{-1})S(\boldsymbol{\gamma} - M^{-1}AS^{-1})^{T}\right] + \lambda \norm{\text{ {\bf vec}}(\boldsymbol{\gamma})}_{0}
		\\
		&= \text{tr}\left[LL^{T}(\boldsymbol{\gamma} - L^{-T}L^{-1}AS^{-1})S(\boldsymbol{\gamma} - L^{-T}L^{-1}AS^{-1})^{T}\right] + \lambda \norm{\text{ {\bf vec}}(\boldsymbol{\gamma})}_{0}
		\\
		&= \text{tr}\left[L^{T}(\boldsymbol{\gamma} - L^{-T}L^{-1}AS^{-1})S(\boldsymbol{\gamma} - L^{-T}L^{-1}AS^{-1})^{T}L\right] + \lambda \norm{\text{ {\bf vec}}(\boldsymbol{\gamma})}_{0}
		\\
		&=\text{tr}\left[(L^{T}\boldsymbol{\gamma} -L^{-1}AQ^{-T}Q^{-1})QQ^{T}((L^{T}\boldsymbol{\gamma} - L^{-1}AQ^{-T}Q^{-1})^{T}\right] + \lambda \norm{\text{ {\bf vec}}(\boldsymbol{\gamma})}_{0}
		\\
		&=\text{tr}\left[(L^{T}\boldsymbol{\gamma}Q - L^{-1}AQ^{-T})(L^{T}\boldsymbol{\gamma}Q - L^{-1}AQ^{-T})^{T}\right] + \lambda \norm{\text{ {\bf vec}}(\boldsymbol{\gamma})}_{0}
		\\
		&= \text{ {\bf vec}}(L^{T}\boldsymbol{\gamma}Q - L^{-1}AQ^{-T})^{T} \text{{\bf vec}}(L^{T}\boldsymbol{\gamma}Q - L^{-1}AQ^{-T}) + \lambda \norm{\text{{\bf vec}}(\boldsymbol{\gamma})}_{0}.
	\end{split}
\end{equation}

The proportionality in line 2 is up to an additive constant with respect to the action variable, $\boldsymbol{\gamma}$.  We arrive at the final utility by distributing the vectorization and rewriting the inner product as a squared {$\ell_{2}$} norm.

\begin{equation}
	\mathcal{L}(\boldsymbol{\gamma}) =  \norm{ \left[Q^{T} \otimes L^{T}\right]\text{\bf vec}(\boldsymbol{\gamma}) - \text{\bf vec}(L^{-1}AQ^{-T})  }_{2}^{2} + \lambda\norm{ \text{\bf vec}(\boldsymbol{\gamma})}_{0}.
\end{equation}The $l_{0}$ norm penalty yields a difficult combinatorial optimization problem even for a relatively small dimensions ($pq \approx 30$).  Thus, one may use an $\ell_{1}$ norm as the most straightforward approximation to the $\ell_{0}$ norm, yielding the loss function: 

\begin{equation}
	\mathcal{L}(\boldsymbol{\gamma}) =  \norm{ \left[Q^{T} \otimes L^{T}\right]\text{\bf vec}(\boldsymbol{\gamma}) - \text{\bf vec}(L^{-1}AQ^{-T})  }_{2}^{2} + \lambda\norm{ \text{\bf vec}(\boldsymbol{\gamma})}_{1}.
\end{equation}

\section{Derivation of the loss function under fixed predictors}

We devote this section to deriving an analogous loss function for multivariate regression when the predictors are assumed fixed.  Notice that this is essentially an extension of \cite{HahnCarvalho} to the multiple response case and adds to the works of \cite{brown1998multivariate} and \cite{wangSUR} by providing a posterior summary strategy that relies on more than just marginal quantities like posterior inclusion probabilities.

Suppose we observe $N$ realizations of the predictor vector defining the design matrix $\textbf{X} \in \mathbb{R}^{Nxp}$.  Future realizations $\tilde{\textbf{Y}} \in \mathbb{R}^{Nxq}$ at this fixed set of predictors are generated from a matrix normal distribution:

\begin{equation}
\mathbf{\tilde{Y}} \sim \textrm{Matrix Normal}_{N,q} \left(\textbf{X} \boldsymbol{\gamma}^{T}, \hspace{1mm} \mathbb{I}_{N x N}, \hspace{1mm} \Psi_{q x q} \right).\label{matrixdist}	
\end{equation}In this case, the optimal posterior summary  $\boldsymbol{\gamma}^*$ minimizes the expected loss $\mathcal{L}_{\lambda}(\boldsymbol{\gamma}) = \mathbb{E}[\mathcal{L}_{\lambda}(\tilde{\bf Y},\Theta,\boldsymbol{\gamma})]$.  Here, the expectation is taken over the joint space of the predictive and posterior distributions: $p(\tilde{\bf Y}, \Theta \vert \textbf{Y}, \textbf{X})$ {\it where $\tilde{X}$ is now absent since we are relegated to predicting at the observed covariate matrix} $\textbf{X}$.  We define the utility function using the negative kernel of distribution (\ref{matrixdist}) where, as before, $\boldsymbol{\gamma}$ is the summary defining the sparsified linear predictor and $\Omega = \Psi^{-1}$:

\begin{equation}
	\begin{split}
	\mathcal{L}_{\lambda}(\tilde{\textbf{Y}},\Theta,\boldsymbol{\gamma}) &= \frac{1}{2}\text{tr}\left[\Omega(\tilde{\textbf{Y}} - \textbf{X}\boldsymbol{\gamma}^{T})^{T} (\tilde{\textbf{Y}} - \textbf{X}\boldsymbol{\gamma}^{T}) \right] + \lambda\norm{\text{{\bf vec}}(\boldsymbol{\gamma})}_{0},
	\end{split}
\end{equation}

Expanding the inner product and dropping terms that do not involve $\boldsymbol{\gamma}$, we define the loss up to proportionality:

\begin{equation}
	\begin{split}
	\mathcal{L}_{\lambda}(\tilde{\textbf{Y}},\Theta,\boldsymbol{\gamma}) &\propto \text{tr}\left[\Omega( \boldsymbol{\gamma}\textbf{X}^{T}\textbf{X}\boldsymbol{\gamma}^{T} - 2\tilde{\textbf{Y}}^{T}\textbf{X}\boldsymbol{\gamma}^{T} ) \right] + \lambda\norm{\text{{\bf vec}}(\boldsymbol{\gamma})}_{0}.
	\end{split}
\end{equation}Analogous to the stochastic predictors derivation, we integrate over $(\tilde{Y},\Theta)$ to obtain our expected loss:

\begin{equation}
	\begin{split}
		\mathcal{L}_{\lambda}(\boldsymbol{\gamma}) &= \mathbb{E}[\mathcal{L}_{\lambda}(\tilde{\bf Y},\Theta,\boldsymbol{\gamma})]
		\\
		&= \text{tr}[M\boldsymbol{\gamma}S_{f}\boldsymbol{\gamma}^{T}] - 2\text{tr}[A_{f}\boldsymbol{\gamma}^{T} ] + \lambda\norm{\text{{\bf vec}}(\boldsymbol{\gamma})}_{0}.
	\end{split}
\end{equation}

where, similar to the random predictor case, $A_{f}=\mathbb{E}[\Omega\tilde{\textbf{Y}}^{T}\textbf{X}]$, $S_{f}=\textbf{X}^{T}\textbf{X}$, $M=\overline{\Omega}$, and the overlines denote posterior means. The subscript $f$ is used to denote quantities calculated at {\it fixed} design points $\textbf{X}$.   Defining the Cholesky decompositions: $M = LL^{T}$ and $S_{f} = Q_{f}Q_{f}^{T}$ and replacing the $\ell_0$ norm with the $\ell_1$ norm, this expression can be formulated in the form of a standard penalized regression problem:

\begin{equation}
	\mathcal{L}_{\lambda}(\boldsymbol{\gamma}) =  \norm{ \left[Q_{f}^{T} \otimes L^{T}\right]\text{\bf vec}(\boldsymbol{\gamma}) - \text{\bf vec}(L^{-1}A_{f}Q_{f}^{-T})  }_{2}^{2} + \lambda\norm{ \text{\bf vec}(\boldsymbol{\gamma})}_{1}\label{lassoformfixed}
\end{equation}with covariates $Q_{f}^{T} \otimes L^{T}$, ``data" $L^{-1}A_{f}Q_{f}^{-T}$, and regression coefficients $\boldsymbol{\gamma}$. Accordingly, (\ref{lasso_form}) can be optimized using existing software such as the {\tt lars} R package of \cite{Efron}.

We use loss function (\ref{lassoformfixed}) as a point of comparison to demonstrate how incorporating covariate uncertainty may impact the summarization procedure in our applications.

%% ** The bibliograhy **
\bibliographystyle{imsart-nameyear.bst}
\bibliography{MultiVSStochasticCovariates}% place <bib-data-file> 

% ** Acknowledgements **
% \begin{acknowledgement}
% \end{acknowledgement}

\end{document}